\documentclass[letterpaper, 10 pt, conference]{ieeeconf}  % Comment this line out if you need a4paper

\IEEEoverridecommandlockouts                              % This command is only needed if 
\usepackage{definitions}

\title{\LARGE \bf
Cooperative Multi-Agent Path Finding: Beyond Path Planning and Collision Avoidance
}

\author{Nir Greshler$^{1}$, Ofir Gordon$^{2}$, Oren Salzman$^{2}$, and Nahum Shimkin$^{1}$% <-this % stops a space
\thanks{$^{1}$Nir Greshler and Nahum Shimkin are with Viterbi Faculty of Electrical \& Computer Engineering, Technion, Haifa, Israel. Emails:
        {\tt\small nirgreshler@campus.technion.ac.il, shimkin@ee.technion.ac.il}}%
\thanks{$^{2}$Ofir Gordon and Oren Salzman are with Department of Computer Science, Technion, Haifa, Israel. Emails:
        {\tt\small ofirgo@campus.technion.ac.il, osalzman@cs.technion.ac.il}}%
}

\begin{document}

\maketitle
\thispagestyle{empty}
\pagestyle{empty}

%%%%%%%%%%%%%%%%%%%%%%%%%%%%%%%%%%%%%%%%%%%%%%%%%%%%%%%%%%%%%%%%%%%%%%%%%%%%%%%%
\begin{abstract}
We introduce the \emph{Cooperative Multi-Agent Path Finding ({\cmapf})} problem, an extension to the classical MAPF problem, where cooperative behavior is incorporated. 
In this setting, a group of autonomous agents operate in a shared environment and have to complete \emph{cooperative tasks} while avoiding collisions with the other agents in the group. 
This extension naturally models many real-world applications, where groups of agents are required to collaborate in order to complete a given task.
To this end, we formalize the {\cmapf} problem and introduce \emph{Cooperative Conflict-Based Search (\ouralg)}, a \cbs-based algorithm for solving the problem optimally for a wide set of {\cmapf} problems.
\ouralg uses a cooperation-planning module integrated into \cbs such that cooperation planning is decoupled from path planning.
Finally, we present empirical results on several MAPF benchmarks demonstrating our algorithm's properties.
\end{abstract}

%%%%%%%%%%%%%%%%%%%%%%%%%%%%%%%%%%%%%%%%%%%%%%%%%%%%%%%%%%%%%%%%%%%%%%%%%%%%%%%%
\section{Introduction and Related Work}
\label{sec:introduction}
The Multi-Agent Path-Finding (MAPF) problem is a special and important type of the more general Multi-Agent Planning (MAP) problem \cite{torreno2017}. 
In MAPF \cite{stern2019}, the task is to find paths for each agent in a group, from a start to a goal location, where interactions between agents are restricted to collision avoidance, as agents move in a shared environment.
While relevant to many real-world applications, such as warehouse automation \cite{wurman2008}, autonomous vehicles \cite{dresner2008,vsvancara2019} and robotics \cite{honig2016}, recent research in the field has focused on expanding the classical MAPF framework to fit more real-world applications \cite{ma2016a,felner2017,salzman2020}.

A main research direction towards the real-world applicability of MAPF problems is the problem of lifelong MAPF, also known as the Multi-Agent Pickup and Delivery (MAPD) problem. 
In this problem, a group of autonomous agents operate in a shared environment to complete a stream of incoming tasks, each with start and goal locations, while avoiding collisions with each others \cite{ma2017,liu2019}. 
A similar problem, studied by Ma et al.~\cite{ma2016b} is the package-exchange robot-routing problem (PERR) where payload exchanges and transfers are allowed thus enabling the modelling of more general transportation problems. 

% \changed{
% The classical MAPF problem is inherently cooperative, since each agent has to arrive at its goal, without colliding with other agents.
% However, in many real-world applications, agents that operate in a shared environment, are often \emph{heterogeneous} \cite{atzmon2020a} and may have a different set of abilities and restrictions.
% For instance, in the domain of smart urban mobility \cite{bucchiarone19}, where the system is composed of heterogeneous autonomous agents, agents collaborate in order to leverage other agents' capabilities to perform tasks more efficiently.
% In this work, we introduce the \emph{Cooperative-MAPF ({\cmapf})} framework, an extension to MAPF, in which a group of agents collaborate towards completing a \emph{cooperative task}.
% Therefore, in the {\cmapf} framework, achieving goals and completing tasks may not depend only on avoiding collisions between agents, but also on actively coordinating their actions.
% Simply put, we may want agents not just to ``not interrupt'' each other, but at the same time help each other achieve their goals.
% We term this a \emph{truly cooperative} setting.}

In this work, we introduce the \emph{Cooperative-MAPF ({\cmapf})} framework, a MAPF extension, in which a group of agents collaborate towards completing a \emph{cooperative task}.
The classical MAPF problem is inherently cooperative, since each agent has to arrive at its goal, without colliding with other agents.
However, in many real-world applications, agents that operate in a shared environment are often \emph{heterogeneous} \cite{atzmon2020a} and may have a different set of abilities and restrictions.
Therefore, in the {\cmapf} framework, achieving goals and completing tasks may not depend only on avoiding collisions between agents, but also on actively coordinating their actions.
Simply put, we may want agents not just to ``not interrupt'' each other, but also help each other achieve their goals.
We term this a \emph{truly cooperative} setting.

\begin{figure}[ht]
    \centering
    \includegraphics[width=1.0\columnwidth]{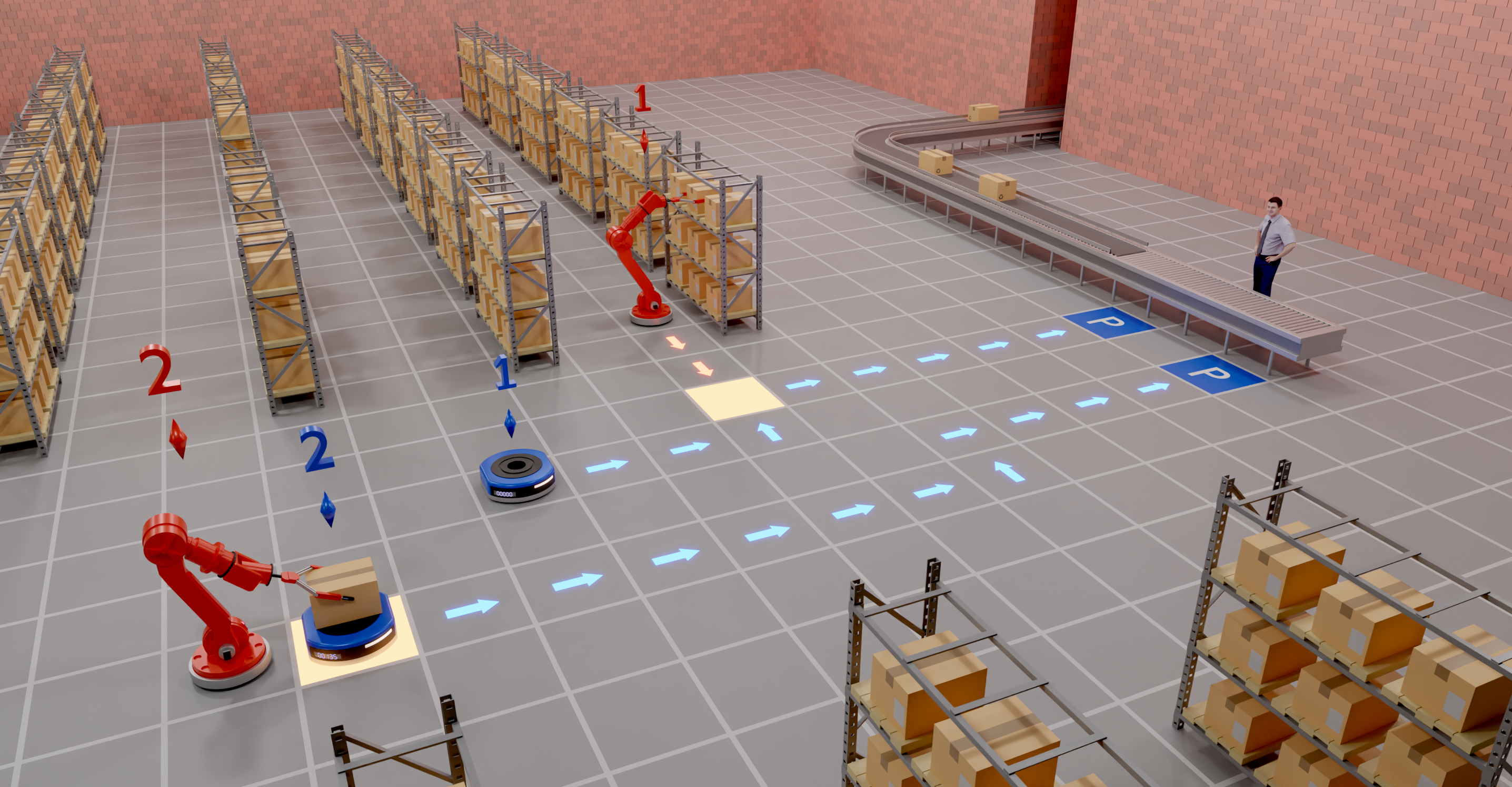}
    \caption{Two pairs of robots operate in a warehouse--two grasp units and two transfer units. 
    Grasp unit $\#1$ arrived at the task start location, i.e., next to the shelf. It will pick up the box and then drive to the meeting location (marked with a yellow square) to transfer the box to transfer unit $\#1$. 
    The transfer unit has a path (marked with blue arrows) to the meeting point, and from there to the task goal (the P square), where the box will be picked by a human employee.
    The second pair of robots ($\#2$) are at their meeting location.}
    \label{fig:warehouse-illustration}
    \vspace{-0.5em}
\end{figure}

Our motivating problem is taken from the warehouse-automation domain \cite{wurman2008}. 
In this problem, storage locations host inventory pods that hold goods of different kinds. 
%A large number of robots operate
Robots operate autonomously in the warehouse, picking up and carrying inventory pods to designated drop-off locations, where goods are manually taken off the pods for packaging. %The robots then carry the pods back to their location in the warehouse.
In this scenario, the robot's main task is to transport the pods around the warehouse, and we refer to robots executing such tasks as \emph{transfer units}.
Research in a different, yet closely-related area, has studied the problem of autonomous robotic arms capable of picking-up a specific item from an inventory pod \cite{correll2016}. 
We refer to a moving robot with such arm as a \emph{grasp unit}.
This motivates the investigation of an improved warehouse scenario, where robots of two types, grasp and transfer units, can work together in coordination (for example, by scheduling a meeting between them) to improve some optimization objective. 
For instance, the number of completed tasks for a given time period.
This motivating example is depicted in Fig. \ref{fig:warehouse-illustration}.

We incorporate a truly cooperative behavior to classical MAPF by assigning cooperative tasks (rather than goals) to agents, similar to (non-cooperative) tasks defined in the MAPD literature~\cite{ma2017,liu2019}.
Agents cooperate in the context of these cooperative tasks, and are only able to complete %their 
tasks by coordinating their actions and goals with each other.

We suggest a formulation to the {\cmapf} problem which is derived from the classical MAPF formulation~\cite{stern2019}. In addition, we discuss differences and further extensions to the {\cmapf} framework which can be used towards achieving more cooperative capabilities in a MAPF problem. 
In the suggested formulation, presented in Section~\ref{sec:background}, there is more than one set of agents, possibly representing heterogeneous real-world agents, and we specifically focus on the case of two sets of agents.
The cooperation between agents is restricted to the form of \emph{meetings}, where agents have to schedule a meeting location and time to complete a task.
We also discuss other forms of agent interactions, and generalizations to the suggested formulation.
Besides the aforementioned warehouse problem, more real-world problems can be modeled using the {\cmapf} framework, such as the involvement of aerial robots in fulfilment centers~\cite{shome2020}, the truck-and-drone ``last-mile'' delivery problem~\cite{murray2020} and multi-drone delivery using transit networks~\cite{choudhury2020}.

Based on the suggested formulation, we introduce (in Section~\ref{sec:algorithm}) \emph{Cooperative Conflict-Based Search (\ouralg)}, an optimal three-level algorithm that is heavily based on two previously-suggested optimal algorithms: 
the well-known Conflict-Based Search (\cbs)~\cite{sharon2015} for solving a classical MAPF problem and the Conflict-Based Search with Optimal Task Assignment (\cbsta)~\cite{honig2018} for solving the anonymous MAPF problem, where we also need to assign goals (or tasks) to each agent.
We define, similarly to MAPD problems~\cite{cap2015,ma2017}, a notion of \emph{well-formed} problem instances, representing realistic and practical environments in MAPF domains, for which a solution to the {\cmapf} problem is guaranteed to exist.
\newtext{Finally, we introduce two improvements to the basic version of \ouralg.}

For clarity of exposition, the description of our \ouralg algorithm is based on the original \cbs algorithm which has numerous extensions and improvements. Many of these improvements can be immediately applied to \ouralg, as we discuss in Section~\ref{sec:discussion}.

A theoretical analysis of \ouralg is presented in Section~\ref{sec:theory} where we prove that \ouralg finds an \emph{optimal} solution for any well-formed {\cmapf} problem instance (formally defined in Section~\ref{sec:well-formed}).
Since the MAPF problem is NP-hard, so is {\cmapf}. 
We therefore discuss \ouralg runtime, provide a qualitative analysis, and show empirically that it can solve nontrivial problem instances. 
More specifically, we present results of running \ouralg on several MAPF benchmarks (detailed in Section~\ref{sec:experiments}).
\newtext{We show that our two suggested \ouralg improvements significantly improve the algorithm's performance.}
% a simulated warehouse problem, using Asprilo \cite{gebser2018} (detailed in Section~\ref{sec:experiments}).

Finally, in Section~\ref{sec:discussion} we discuss some extensions and research directions, specifically for \ouralg, but more importantly, general for the {\cmapf} framework.

% %%%%%%%%%%%%%%%%%%%%%%%%%%%%%%%%%%%%%%%%%%%%%%%%%%%%%%%%%%%%%%%%%%%%%%%%%%%%%%%%
\section{Background and Setting}
\label{sec:background}
We first describe and formulate the classical MAPF problem followed by a formulation of our proposed Cooperative-MAPF ({\cmapf}) framework. Then, we define the objective function used in {\cmapf}.

\subsection{Classical MAPF}
In the classical MAPF problem \cite{stern2019}, we are given an undirected graph $G=\argument{V,E}$ whose vertices $V$ correspond to locations and whose edges $E$ correspond to connections between the locations that the agents can move along.~${A=\set{a_1,\dots,a_\numagents}}$ is a set of $\numagents$ agents, each is provided with a start and goal location, $\argument{s_i, g_i}$ s.t.~${s_i,g_i\in V}$.

Time is discretized and at each time step, each agent can either \textit{move} on the graph or \textit{wait} at its current vertex. A feasible MAPF solution is a set of paths $\calP=\set{p_1,\dots,p_k}$ such that $p_i$ is a path for agent $a_i$ from vertex $s_i$ to vertex $g_i$ and there are no conflicts between any two paths in~$\calP$. We consider two types of conflicts---a \emph{vertex conflict}, in which two agents occupy the same vertex at the same time step, and an \emph{edge conflict} (or \emph{swapping conflict}), in which two agents traverse the same edge from opposite sides (``switch sides'') at the same time step. An optimal solution is a feasible set of paths~$\calP$ which optimizes some objective function (specifically defined in Section~\ref{sec:objective}).

\subsection{Cooperative-MAPF ({\cmapf})} 
\label{sec:cooperative-mapf}
We wish to incorporate cooperative behavior into the classical MAPF problem. 
This is done by replacing agent goals with a set of \emph{cooperative tasks}, i.e., tasks that require the cooperation and coordination of a group of agents in order to be completed. 
Specifically, here we limit ourselves to cooperative tasks (simply referred to as tasks in the rest of this paper) that require pre-defined pairs of agents to work together. 
We discuss possible extensions in Section~\ref{sec:discussion}.

In the {\cmapf} problem we are given an undirected graph $G=\argument{V, E}$.
The set of agents $A$ consists of two distinguishable sets, i.e.,~${A=\groupA \cup \groupB}$.
Each set includes~$\numagents$ agents of a specific type, namely~${\groupA=\set{\agentAi{1},\dots,\agentAi{\numagents}}}$ and~${\groupB=\set{\agentBi{1},\dots,\agentBi{\numagents}}}$. 
The two types of agents may differ in their traversal capabilities or possible actions in a location (for instance, picking up an object). 
We are also given a set of tasks~${\taskset=\set{\task{1},\dots,\task{\numagents}}}$ s.t. each task~$\task{i}$ is assigned to a pair of agents~${\argument{\agentAi{i}, \agentBi{i}}}$. 
We refer to~${\agentAi{i}}$ and~${\agentBi{i}}$ as the \emph{\agentAname} and \emph{\agentBname} agents, respectively.
Each task~${\task{i} \in \taskset}$ is defined by a start location $s_i$ and a goal location $g_i$.

Each agent has a unique start location given by a function~${\agentstart:A \rightarrow V}$ s.t.~${\agentstartfunc{a}}$ is the location of agent~$a$ at time step 0. 
An agent goal is not directly given but rather derived from its assigned task. 
In our setting, a task~${\task{i} = (s_i, g_i)}$ for agents~${(\agentAi{i}, \agentBi{i})}$ is composed of the following steps: 
(i)~moving the {\agentAname} agent $\agentAi{i}$ to the task's start location~$s_i$,
(ii)~moving both agents to a so-called \emph{meeting}~${m_i=\meeting{i}}$ where~${\meetinglocation{i} \in V}$ is the meeting location and~${\meetingtime{i}}$ is the meeting time step, both of which are computed by the algorithm (and not specified by the task\footnote{Note that a meeting $m_i$ is defined by its location and time. Thus, when referring to a meeting, we mean both.}),
(iii)~moving the {\agentBname} agent to the task's goal location $g_i$.
For a visualization, see Fig. \ref{fig:warehouse-illustration}.

% More specifically, for a task to be completed, the {\agentAname} agent has to plan a path from its start location to the task start location and then to \emph{meet} with the {\agentBname} agent. 
% The {\agentBname} has to plan a path from its start location to first meet the {\agentAname} and then complete the task by planning a path to the task's goal location. 
% We denote a \emph{meeting} between agents~${\argument{\agentAi{i}, \agentBi{i}}}$ by~${\meeting{i}}$ where $\meetinglocation{i} \in V$ is the meeting location and $\meetingtime{i}$ is the meeting time step.

Formally, a solution to a {\cmapf} instance is a set of paths pairs~${\calP=\set{(p_1^\agentA, p_1^\agentB),\dots,(p_k^\agentA, p_k^\agentB)}}$ s.t. for each pair~${1 \leq i \leq k}, \:$ $p_i^\agentA, p_i^\agentB$ start in~${\agentstartfunc{\agentAi{i}}}$ and $\agentstartfunc{\agentBi{i}}$, respectively. Path~${p_i^\agentA}$ goes through~${s_i}$ at some time step $t_i$, and both paths contain a meeting at vertex~${\meetinglocation{i}}$ at the same time~${\meetingtime{i}}$ s.t.~${t_i \leq \meetingtime{i}}$. 
Finally,~${p_i^\agentA}$ ends in vertex~${\meetinglocation{i}}$ at time~${\meetingtime{i}}$~and ${p_i^\agentB}$ ends in vertex~${g_i}$. 
Similarly to classical MAPF, in order for a solution to be feasible, there should be no conflicts between the paths in $\mathcal{P}$, with the exception that the paths of agents sharing a task intersect at their meeting point.

\subsection{Objective functions for Cooperative MAPF}
\label{sec:objective}
Arguably, the most common objective functions used in classical MAPF to evaluate solutions are \emph{makespan (MKSP)} and \emph{sum-of-costs (SOC)} \cite{stern2019}, both to be minimized.
\newtext{
MKSP is defined as the number of time steps required for all agents to reach their target, while SOC is the sum of time steps required by each agent to complete all tasks.
In this paper we focus on the SOC objective, which is,
arguably, more natural for our setting---it implicitly minimizes both the time it takes to complete a task, and the time the {\agentAname} finishes its part in the task.
We note that all results presented can be applied to the MKSP objective as well.} 
% We similarly define the corresponding SOC objective function for the cooperative case.
The sum of costs of~$\calP$ is defined as $\sum_{1\leq i \leq k}{\setsize{p_i^\agentA}+|p_i^\agentB|}$. Wait actions are counted until an agent finishes its plan (i.e., after the meeting for $\agentAi{i}$ and after arriving at $g_i$ for $\agentBi{i}$).
% \ask{TODO: we can probably omit the SOC superscript from everywhere}
% \begin{itemize}
%     \item \removed{\textbf{Makespan}. The number of time steps required for all tasks to be completed. For a solution~${\calP=\set{(p_1^\agentA, p_1^\agentB),\dots,(p_k^\agentA, p_k^\agentB)}}$ the makespan of~$\calP$ is defined as~${\max_{1\leq i \leq \numagents}{|p_i^\agentB|}}$. Note that we take maximum over paths of the executor%{\agentBname}
%     agent only since~${\forall i, \: |p_i^\agentB| \geq \setsize{p_i^\agentA}}$.}

%     \item \textbf{Sum of costs}. The sum of time steps required by each agent, to complete all tasks. The sum of costs of $\calP$ is defined as $\sum_{1\leq i \leq k}{\setsize{p_i^\agentA}+|p_i^\agentB|}$. Wait actions are counted until an agent finishes its plan (i.e., after the meeting for $\agentAi{i}$ and after arriving at $g_i$ for $\agentBi{i}$).
% \end{itemize}

\subsection{Well-Formed {\cmapf} Instances}
\label{sec:well-formed}

It is possible to efficiently check if a MAPF instance is solvable \cite{yu2014}. 
However, checking if a {\cmapf} instance is solvable is not trivial due to the additional requirement that meetings need to be computed.
Therefore, we restrict our discussion to \emph{well-formed} instances \cite{cap2015, ma2017}. 
The intuition behind the well-formed definition is that agents can rest (that is, stay forever) in locations, called \emph{endpoints}, where they cannot block the execution of other tasks.

% \OG{Since we're omitting the proofs, I think that we can explain the intuition of well-formed without using the daunting formulation. I'm writing an alternative suggestions, let me know what you think}.
% \ask{TODO: we removed all notations of well-formed. Make sure we're not using them later.}

The set~$V_{ep}$ of endpoints contains the start locations of all agents together with the start and goal locations of all tasks.
The complement set~${V \setminus V_{ep}}$ contains all non-endpoints vertices.
We define a pair of vertices as \emph{connected} if there exists a path between them which only includes non-endpoint vertices.

%The set~$V_{ep}$ of endpoints contains the start locations of all agents together with the start and goal locations of all tasks, namely,~${V_{ep}=V_0 \cup \set{s_i}_{i=1}^k \cup \set{g_i}_{i=1}^k}$ where~${V_0=\set{\agentstartfunc{\agentAi{i}}}_{i=1}^\numagents \cup \set{\agentstartfunc{\agentBi{i}}}_{i=1}^\numagents}$.

%The set~${V \setminus V_{ep}}$ contains all non-endpoints vertices.
%We also define a pair of nodes as \emph{connected} if there exists a path from $u$ to $v$ such that all its nodes (besides $u$ and $v$) are in~${V \setminus V_{ep}}$, and denote this property by $\connected{u}{v}$.

\begin{definition}
\label{def:well-formed}
A {\cmapf} instance is well-formed iff
\renewcommand{\labelenumi}{\textbf{(C\arabic{enumi})}}
\begin{enumerate}
    \item 
    \label{C1}
    \changed{
    For every task there exists a vertex~${v
    \in V \setminus V_{ep}}$ that is: (i)~connected with the task start vertex, (ii)~connected with the task goal vertex, and (iii)~connected with the {\agentBname} agent's start vertex.}
    %Namely, 
    % \begin{equation*}
    %     \begin{split}
    %         \text{Namely, }
    %         \forall \task{i} \in \taskset,& \exists u \in V \setminus V_{ep} \text{  s.t. } \\
    %         & {\connected{s_i}{u}},~{\connected{u}{g_i}} \text{ and }{\connected{\agentstartfunc{\agentBi{i}}}{u}}
    %     \end{split}
    % \end{equation*}
    \item 
    \label{C2}
    \changed{
    For every task, the task start vertex is connected with the {\agentAname} agent's start vertex.}
    % Namely,~$\connected{\agentstartfunc{\agentAi{i}}}{s_i}$. 
%    \item $\forall \task{i} \in \taskset, \exists u \in V \setminus V_{ep}$ s.t.~${\connected{s_i}{u}}$,~${\connected{u}{g_i}}$ and~${\connected{\agentstartfunc{\agentBi{i}}}{u}}$.
%    Namely, there exists a free vertex that is: (i) connected with the task start vertex, (ii) connected with the task goal vertex, and (iii)~connected with the {\agentBname} agent's start vertex.
    
 %   \item $\connected{\agentstartfunc{\agentAi{i}}}{s_i}$. 
%    Namely, the task start vertex is connected with the {\agentAname} agent's start vertex.
\end{enumerate}
\end{definition}

\begin{figure}[ht]
    \centering
    \begin{subfigure}{0.22\textwidth}
        \centering
        \begin{tikzpicture}[thick,scale=0.8, every node/.style={scale=0.8}]
        \filldraw[black] (1,0) rectangle (2,1);
        \filldraw[black] (1,1) rectangle (2,2);
        \filldraw[color=blue, fill=none, opacity=1.0, dashed] (1.5,2.5) circle (0.4);
        \filldraw[color=red, fill=none, opacity=1.0, dashed] (0.5,2.5) circle (0.4);
        \filldraw[color=black, fill=red!60] (0.5,0.5) circle (0.4);
        \filldraw[color=black, fill=blue!60] (2.5,0.5) circle (0.4);
        
        \draw[step=1.0,black,thin] (0,0) grid (3,3);
        
        \node at (0.5,0.5) {$\agentAi{1}$};
        \node at (2.5,0.5) {$\agentBi{1}$};
        \node at (0.5,2.5) {$s_1$};
        \node at (1.5,2.5) {$g_1$};
        \end{tikzpicture}
        % \caption{A not well-formed instance.}
        \caption{}
        \label{fig:not-well-formed}
    \end{subfigure}
    \begin{subfigure}{0.22\textwidth}
        \centering
        \begin{tikzpicture}[thick,scale=0.8, every node/.style={scale=0.8}]
        \filldraw[black] (1,0) rectangle (2,1);
        \filldraw[black] (1,2) rectangle (2,3);
        \filldraw[color=blue, fill=none, opacity=1.0, dashed] (2.5,2.5) circle (0.4);
        \filldraw[color=red, fill=none, opacity=1.0, dashed] (0.5,2.5) circle (0.4);
        \filldraw[color=black, fill=red!60] (0.5,0.5) circle (0.4);
        \filldraw[color=black, fill=blue!60] (2.5,0.5) circle (0.4);
        
        \draw[step=1.0,black,thin] (0,0) grid (3,3);
        
        \node at (0.5,0.5) {$\agentAi{1}$};
        \node at (2.5,0.5) {$\agentBi{1}$};
        \node at (0.5,2.5) {$s_1$};
        \node at (2.5,2.5) {$g_1$};
        \end{tikzpicture}
        % \caption{A well-formed instance.}
        \caption{}
        \label{fig:well-formed}
    \end{subfigure}
    \caption{A not well-formed (a) and well-formed (b) {\cmapf} instances. 
    Black squares are obstacles. In (a), no meeting point that adheres to~(C\ref{C1}) in Definition \ref{def:well-formed} exists. In (b), agents can meet at any of the white cells.}
    %\vspace{-1.0em}
\end{figure}
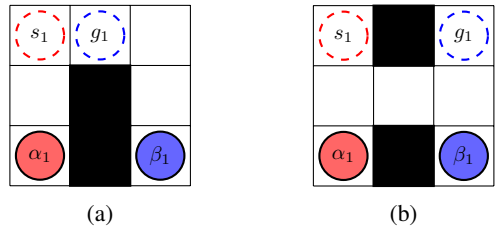

% \OG{consider removing the well-formed figure}
Fig. \ref{fig:not-well-formed} shows an example of a {\cmapf} instance which is not well-formed.
In this problem,~\textbf{(C\ref{C1})} in Definition~\ref{def:well-formed} is violated: there does not exist a vertex~$u$ which is connected to both $s_1$ and $\agentstartfunc{\agentBi{1}}$.
%such that both~$\connected{s_1}{u}$ and~${\connected{\agentstartfunc{\agentBi{1}}}{u}}$ hold.
Fig. \ref{fig:well-formed} shows a well-formed instance: all white squares are valid meeting points.
%\vspace{-0.2em}

\newtext{
We restrict the discussion on Co-MAPF only to well-formed instances.
It allows us to efficiently test if an instance is solvable and ensure completeness of our suggested algorithm. 
% Using the aforementioned definition of well-formed instances, we can restrict the Co-MAPF problem to instances which can be tested for completeness efficiently. 
We summarize this guarantee in the following claim and lemma. Proofs omitted due to space considerations.
}
\begin{claim} \label{claim:well-formed}
Checking if a {\cmapf} instance is well-formed can be done in polynomial time.
\end{claim}

% \newtext{
% \begin{proof}
% Omitted due to space considerations.
% \end{proof}
% }

% \begin{proof}
% \removed{
% Given $V \setminus V_{ep}$, we can find a feasible vertex~$u$ which holds the well-formed requirements for each task~${\task{i}\in{\taskset}}$ by iterating over all vertices in the graph and checking if~${\connected{s_i}{u}}$,~${\connected{u}{g_i}}$ and~${\connected{\agentstartfunc{\agentBi{i}}}{u}}$. 
% In other words, for each potential meeting vertex $u$ we need to check connectivity exactly three times, which can be done in polynomial time in the size of the graph using any graph-search algorithm such as Dijkstra.%{\dij}. 
% Therefore, by performing this test for each task independently, we can answer that the instance is well-formed (if such feasible node found) or conclude that it's not (if none of the vertices are feasible or there is no valid path from the initiator starting point to $s_i$) in polynomial time.
% More specifically, by running Dijkstra%{\dij}
% 's algorithm three times for each task, we get that the complexity of checking the well-formed property is~${\calO\argument{\setsize{\taskset} \cdot\argument{|V|log|V|+|E|}}}$.}
% \end{proof}

\begin{lemma} \label{lemma:well-formed}
Every well-formed {\cmapf} instance is solvable.
% Well-formed {\cmapf} instances are solvable.
\end{lemma}

\section{Cooperative Conflict-Based Search}
\label{sec:algorithm}
We now present the Cooperative Conflict-Based Search (\ouralg) algorithm, a three-level optimal planning algorithm for solving well-formed {\cmapf} problem instances. 
As our suggested algorithm is based on \cbs \cite{sharon2015}, we start with a brief description of it. 
\cbs is a two-level search algorithm.
The high-level performs a best-first search over a so-called conflicts-tree (CT). Each CT node consists of a solution, its cost and a set of constraints.
\cbs finds conflicts in the solution and resolves them by imposing constraints on agents.
A constraint is either a vertex constraint~$\argument{a,v,t}$, or an edge constraint~$\argument{a, u, v, t}$.
The low-level constructs paths for each individual agent while satisfying the imposed constraints.
\cbs resolves conflicts by splitting a CT node and introducing an additional constraint for each agent participating in the conflict at the lower level. 

% \cbs is a two-level search algorithm: the high level searches in the conflicts space (i.e., all possible combinations of conflicts), where the low level searches in the paths space (i.e., all possible combinations of paths, one per agent).
% \cbs works as follows---first it finds an optimal path for each agent independently using some single-agent search algorithm like \astar. \cbs then works to resolve conflicts that occur in the solution: in the high-level search it performs a best-first search upon a constructed conflicts-tree (CT). 
% Any node in the CT consists of a solution, a solution's cost and a set of constraints imposed on agents due to conflicts in the solution. 
% A constraint is either a vertex constraint (due to a vertex conflict) of the form~$\argument{a,v,t}$, which prohibits agent $a$ from being at vertex~$v$ at time step $t$, or an edge constraint (due to an edge conflict) for the form~$\argument{a, u, v, t}$, which prohibits agent $a$ from crossing the edge $(u,v)$ at time step $t$.
% After applying a constraint on a CT node, \cbs runs a low-level search to construct a new solution while satisfying the constraints.

\begin{figure*}[!ht]
    \centering
    \begin{subfigure}{0.34\textwidth}
        \centering
        \begin{tikzpicture}[thick,scale=1.0, every node/.style={scale=1.0}]

% agents
\filldraw[color=black, fill=red!60] (3.5,4-0.5) circle (0.4);
\filldraw[color=black, fill=blue!60] (0.5,4-1.5) circle (0.4);
\filldraw[color=black, fill=red!60] (3.5,4-2.5) circle (0.4);
\filldraw[color=black, fill=blue!60] (2.5,4-3.5) circle (0.4);

\node at (3.5,4-0.5) {$\agentAi{1}$};
\node at (0.5,4-1.5) {$\agentBi{1}$};
\node at (3.5,4-2.5) {$\agentAi{2}$};
\node at (2.5,4-3.5) {$\agentBi{2}$};

% tasks
\filldraw[color=red, fill=none, opacity=1.0, dashed] (1.5,4-0.5) circle (0.4);
\filldraw[color=red, fill=none, opacity=1.0, dashed] (2.5,4-0.5) circle (0.4);
\filldraw[color=blue, fill=none, opacity=1.0, dashed] (1.5,4-1.5) circle (0.4);
\filldraw[color=blue, fill=none, opacity=1.0, dashed] (0.5,4-2.5) circle (0.4);

% meetings
% \node at (1.5, 2.5) [fill=yellow!30,star,star points=5,inner sep=-1.5pt,draw,opacity=0.75,thin] {$m_1$};
% \node at (1.5, 3.5) [fill=yellow!30,star,inner sep=-1.5pt,draw,opacity=0.75,thin] {$m_2$};

\node at (1.5,4-0.5) {$s_1$};
\node at (2.5,4-0.5) {$g_1$};
\node at (1.5,4-1.5) {$s_2$};
\node at (0.5,4-2.5) {$g_2$};

% obstacles
% \filldraw[color=black, fill=black, opacity=1.0, solid] (1.5,2.5) square (1);
\fill [black] (1,1) rectangle (2,2);

% grid
\draw[step=1.0,black,thin] (0,0) grid (4,4);
\foreach \xtick in {0,...,3} {\pgfmathsetmacro\result{\xtick} \node at (\xtick+0.5,-0.3) {\pgfmathprintnumber{\result}}; }
\foreach \ytick in {0,...,3} {\pgfmathsetmacro\result{\ytick} \node at (-.3,3-\ytick+0.5) {\pgfmathprintnumber{\result}}; }

\end{tikzpicture}
% \begin{tikzpicture}[thick,scale=0.7, every node/.style={scale=0.8}]

% % agents
% \filldraw[color=black, fill=red!60] (0.5,0.5) circle (0.4);
% \filldraw[color=black, fill=blue!60] (1.5,0.5) circle (0.4);
% \filldraw[color=black, fill=red!60] (3.5,0.5) circle (0.4);
% \filldraw[color=black, fill=blue!60] (2.5,0.5) circle (0.4);

% \node at (0.5,0.5) {$\agentAi{1}$};
% \node at (1.5,0.5) {$\agentBi{1}$};
% \node at (3.5,0.5) {$\agentAi{2}$};
% \node at (2.5,0.5) {$\agentBi{2}$};

% % tasks
% \filldraw[color=red, fill=none, opacity=1.0, dashed] (1.5,1.5) circle (0.4);
% \filldraw[color=red, fill=none, opacity=1.0, dashed] (2.5,1.5) circle (0.4);
% \filldraw[color=blue, fill=none, opacity=1.0, dashed] (3.5,3.5) circle (0.4);
% \filldraw[color=blue, fill=none, opacity=1.0, dashed] (3.5,2.5) circle (0.4);

% % meetings
% \node at (1.5, 2.5) [fill=yellow!30,star,star points=5,inner sep=-1.5pt,draw,opacity=0.75,thin] {$m_1$};
% \node at (1.5, 3.5) [fill=yellow!30,star,inner sep=-1.5pt,draw,opacity=0.75,thin] {$m_2$};

% \node at (1.5,1.5) {$s_1$};
% \node at (2.5,1.5) {$s_2$};
% \node at (3.5,3.5) {$g_2$};
% \node at (3.5,2.5) {$g_1$};

% % grid
% \draw[step=1.0,black,thin] (0,0) grid (4,4);
% \foreach \xtick in {0,...,3} {\pgfmathsetmacro\result{\xtick} \node at (\xtick+0.5,-0.3) {\pgfmathprintnumber{\result}}; }
% \foreach \ytick in {0,...,3} {\pgfmathsetmacro\result{\ytick} \node at (-.3,\ytick+0.5) {\pgfmathprintnumber{\result}}; }

% \end{tikzpicture}
        \caption{Example instance with two tasks:~$\agentAi{1}$ and~$\agentBi{1}$ execute~$\task{1}$ from~$s_1$ to~$g_1$ and~$\agentAi{2}$ and~$\agentBi{2}$ execute~$\task{2}$ from~$s_2$ to~$g_2$.}
        % The meeting locations in the solution are marked as~$m_1$ and $m_2$ stars.}
        \label{fig:example}
    \end{subfigure}
    \begin{subfigure}{0.65\textwidth}
        \includegraphics[width=0.9\textwidth]{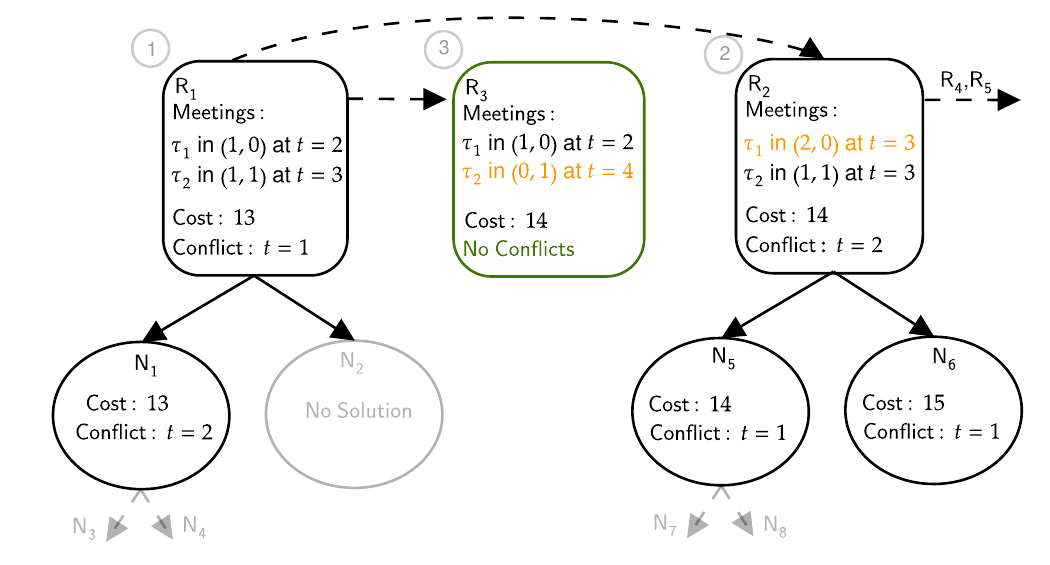}
        \caption{\ouralg search forest.}
        \label{fig:co-cbs-illustration}
    \end{subfigure}
    % \setlength{\abovecaptionskip}{4pt}
  
    % \caption{
    % % \OG{make the text shorter (more concise) (also consider changing the light green to darker green and maybe the arrows to blue)}
    % Example instance with two tasks (left) and a demonstration of \ouralg execution on the presented instance (right).
    % % \caption{Illustration of \ouralg's execution on the scenario depicted in Figure \ref{fig:example}. 
    % Root and regular nodes are denoted with~$R$ and $N$, respectively.
    % The initial root node~$R_1$ contains optimal meetings for $\task{1}, \task{2}$ at $\argument{1,0}$ and $\argument{1,1}$ at times $2$ and $3$, respectively.~$R_1$ has a cost of $13$ and a conflict at time~$t=1$.~$R_1$ is expanded into two new root nodes $R_2$ and $R_3$, where one meeting is replaced in each.~$R_1$ is also split on its first conflict, creating~$N_1, N_2$.
    % Note that $N_2$ has no solution due to the added constraint making the selected meeting infeasible.
    % Regular node $N_1$ is selected next and expanded to $N_3, N_4$, both with no solution.
    % Then~$R_2$ is selected next from \textsc{Open} and expanded to $R_4, R_5, N_5, N_6$.
    % Next, $N_5$ is expanded, creating two more nodes with no solution.
    % Finally, $R_3$ is selected. 
    % Since it has a cost similar to the optimal cost, and no conflicts, it is returned as a feasible optimal solution.}
    
    \caption{
    Example execution of \ouralg (\subref{fig:co-cbs-illustration}) on the instance depicted in (\subref{fig:example}).
    Root and regular nodes are denoted with~$R$ and~$N$, respectively.
    The initial root node $R_1$ contains the optimal set of meetings, has a cost of $13$ and a conflict at $t=1$.
    It is expanded to root nodes $R_2, R_3$ (with cost $14$) and split on its first conflict, creating $N_1, N_2$.
    $N_2$ has no solution, thus $N_1$ will be chosen next for expansion, creating $N_3, N_4$, both with no solution.
    In the next iterations, \ouralg will expand $R_2$ (to $R_4, R_5, N_5, N_6$), then $N_5$, and finally $R_3$.
    Since $R_3$ has no conflicts, it will be returned as a feasible optimal solution.
    }
    
    \label{fig:co-cbs-example}
    % each with a different set of meetings.
    % Each blue node represents a root of a CT with a defined set of meetings. The dashed-green arrows represent creating new root nodes during a previous root node expansion. Each new root node calls for the meetings planner to get the next best meeting. Orange arrows represent splits in a CT to resolve conflict, similar to \cbs.\NGR{finish caption}
    %\vspace{-0.8em}
\end{figure*}
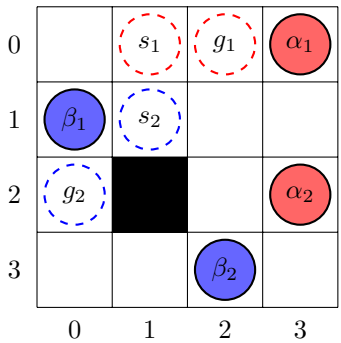
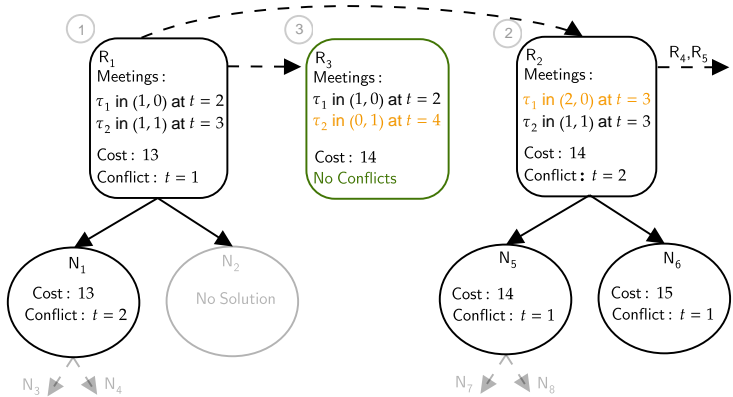

We now continue with an overview of \ouralg (depicted in Fig. \ref{fig:co-cbs-example} and outlined in  Algorithm~\ref{alg:co-cbs}). We then continue with lower-level details.
\subsubsection{Algorithm overview}
\ouralg is a search algorithm based on \cbs that considers the cooperative aspect of the problem.
More specifically, \ouralg consists of three levels of search in three different spaces (similar to~\cite{honig2018} and~\cite{surynek2020}):
(i)~the \emph{meetings space}, 
(ii)~the \emph{conflicts space} 
and 
(iii)~the \emph{paths space}.
The meetings space contains all possible combinations of meetings, one for each task.
We'll refer to the three levels of search as the meetings level, conflicts level and paths level, respectively.

\begin{algorithm}[ht]
\caption{Cooperative Conflict-Based Search (\ouralg)}
\label{alg:co-cbs}
\begin{algorithmic}[1]
\State \textbf{Input:} $G,\groupA,\groupB,\agentstart,\taskset$ \Comment{{\cmapf} problem instance}
\State \textbf{Returns:} optimal path for each agent

\ForAll{$\task{i} \in \taskset$} \Comment{using Algorithm \ref{alg:meetings_table} for each task}
    \State $\meetingtable{i}{} \gets compute\_meetings\_table(\task{i},\agentAi{i},\agentBi{i})$ 
\EndFor

\State $R.constraints \gets \emptyset$
\State $R.meetings \gets $ get the initial set of optimal meetings
\State $R.root \gets $ True
\State $R.solution \gets plan\_paths()$ \Comment{to and from meetings}
\State $R.cost \gets compute\_cost(R.solution)$
\State insert $R$ to \textsc{OpenRoots}
\While{\textsc{Open} $not$ $empty$ or \textsc{OpenRoots} $not$ $empty$} 
    \State{$N \gets$ lowest cost node from \textsc{Open}$\cup$\textsc{OpenRoots}} %\Comment{lowest solution cost} 
    \State Validate the path in $N$ until a conflict occurs
    \If{$N$ has no conflicts}
        \State \Return $N.solution$ \Comment{$N$ is goal}
    \EndIf
    \If{$N.root\ is\ True$}
        \State{$expand\_root(N)$} \Comment{using Algorithm \ref{alg:expand_root}}
    \EndIf
    \State $C\gets$ first conflict $\argument{a_i, a_j, v, t}$ in $N$
    
    \ForAll{$agent\:a_i\:in\:C$} 
        \State $A\gets$ new node
        \State $A.constraints \gets N.constraints + \argument{a_i, v, t}$
        \State $A.meetings \gets N.meetings$
        \State $A.root \gets$  False
        \State $A.solution \gets N.solution$
        \State Update $A.solution$ by invoking $plan\_paths(a_i)$
        \State $A.cost \gets compute\_cost(A.solution)$
        \State Insert $A$ to \textsc{Open}
    \EndFor
    
\EndWhile

\end{algorithmic}
\end{algorithm}

\ouralg simultaneously searches over all possible meetings and for each meeting, over all possible paths. 
To perform this search in a systematic and efficient manner, we need to consider an \emph{ordering} of the meetings. 
% \OG{I think that the next sentence can be removed (Oren said to rephrase it by I think that anyway it is more confuses than it helps)}
% This is analogous to the way conflicts induce an ordering sets of paths and are used by \cbs. 
Indeed, in Equation~\ref{eq:meeting_cost_soc} we define a meeting's cost which is dependent both on the meeting's location and time.
To efficiently traverse the set of possible meetings, we introduce the notion of a \emph{Meetings Table} which stores for each meeting location the currently-best meeting time.
As we will see, this table will allow us to iterate over all meetings in a best-first manner.

In contrast to \cbs that constructs a single conflicts-tree (CT), \ouralg creates a forest of CTs, similar to~\cite{honig2018}. 
Each CT starts in a \emph{root} node and corresponds to a specific set of meetings (a specific meeting for each task).
In \ouralg, each CT node has two additional fields (when compared to \cbs): \emph{root} specifies if the node is a root or a \emph{regular} node and \emph{meetings} specifies the current set of meetings (one for each task) which is used during the path-level search.

\ouralg starts with a single root node, with the optimal set of meetings %(defined in Equation \ref{eq:optimal_meetings_set}), 
(see Equation \ref{eq:optimal_meetings_set}), while ignoring possible conflicts between agents.
In each iteration, \ouralg selects a lowest-cost node from the \textsc{Open} list (either a root or regular node), in a best-first approach similar to \cbs.
Whenever a root node is selected, in addition to splitting the tree due to a conflict, %in the solution, 
\ouralg also expands it in the meetings space by generating the next best sets of meetings. 
Namely, new root nodes are created only on demand.
For each expanded node, given its set of meetings and constraints, the paths level computes a solution by planning the different steps a task solution is composed of (%see
Section~\ref{sec:cooperative-mapf}).

% \begin{algorithm*}[ht]
% \caption{Calculate Meetings Table}
% \label{alg:meetings_table}
% \begin{algorithmic}[1]
% \State \textbf{Input:} A Task $\task{i}$ and two assigned agents
% \State \textbf{Returns:} $\meetingtable{i}$, the meetings table for task $\task{i}$

% \State Plan from {\agentAname} agent start vertex to task start vertex to compute $d\argument{\agentstartfunc{\agentA{i}}, s_i}$ \Comment{using \astar}

% \State Plan from task start vertex to all other vertices in one pass to compute $d\argument{s_i, v}, \forall v \in V$ \Comment{using \dij}

% \State Plan from {\agentBname} agent start vertex to all other vertices in one pass to compute $d\argument{\agentstartfunc{\agentB{i}}, v}, \forall v \in V$ \Comment{using \dij}

% \State Plan from task goal vertex to all other vertices in one pass to compute $d\argument{v, g_i}, \forall v \in V$ \Comment{using \dij}

% \ForAll{$v \in V$}
%     \State Calculate earliest possible meeting time in $v$, $\meetingtime{i}\argument{v}$
%     \State Calculate $\meetingtable{i}\argument{v}$, the cost of meeting in $v$, i.e., the cost (MKSP or SOC) to complete the task with a meeting in $v$
%     \State Add a row to the table with $\argument{v, \meetingtable{i}\argument{v}, \meetingtime{i}\argument{v}}$ \Comment{for each vertex, store meeting cost and earliest meeting time}
% \EndFor
% \State Cache the table for later use

% \end{algorithmic}
% \end{algorithm*}
\begin{algorithm}[ht]
\caption{Compute Meetings Table}
\label{alg:meetings_table}
\begin{algorithmic}[1]
\State \textbf{Input:} A Task $\task{i}$ and two assigned agents
\State \textbf{Returns:} $\meetingtable{i}{}$, the meetings table for task $\task{i}$

\State Compute $d\argument{\agentstartfunc{\agentAi{i}}, s_i}$ \Comment{{\agentAname} start to task start}%\footnotemark

\State Compute $d\argument{s_i, v}, \forall v \in V$ \Comment{task start to all}

\State Compute $d\argument{\agentstartfunc{\agentBi{i}}, v}, \forall v \in V$ \Comment{{\agentBname} start to all}

\State Compute $d\argument{v, g_i}, \forall v \in V$ \Comment{task goal to all}

\ForAll{$v \in V$}
    \State Calculate $\earliestvertextime{i}$ \Comment{earliest meeting time in $v$}
    \State Calculate $\meetingtable{i}{}\argument{v}$ \Comment{task cost with meeting at $v$}
    \State Store $\argument{v, \earliestvertextime{i}, \meetingtable{i}{}\argument{v}}$ in table %\Comment{meeting location, earliest time and cost}
\EndFor
% \State Cache the table for later use

\end{algorithmic}
\end{algorithm}
\setlength{\textfloatsep}{1em}

\begin{algorithm}[ht]
\caption{Expand root}
\label{alg:expand_root}
\begin{algorithmic}[1]
\State \textbf{Input:} Meetings tables of all tasks, a root node $P$

\ForAll{$\task{i} \in \taskset$} \Comment{loop over all tasks} 
    \State $R = $ new node
    \State $R.constraints = \emptyset$

    \State $R.meetings = P.meetings$

    \State $R.meetings[\task{i}] = get\_next\_meeting(\meetingtable{i}{})$ %\Comment{change \textbf{only} $\task{i}$'s meeting}
    
    \State $R.root$ = True
    \State Update $R.solution$ by invoking $plan\_paths(\agentAi{i}, \agentBi{i})$
    \State $R.cost=compute\_cost(R.solution)$
    \State insert $R$ to \textsc{OpenRoots} %\Comment{we split the tree by creating a new node for every task}
\EndFor

\end{algorithmic}
\end{algorithm}

\subsubsection{Computing the Meetings Table}
\label{sec:meeting_tables}
We denote the cost of a meeting $m_i=\meeting{i}$ as $C_i(\meetinglocation{i},\meetingtime{i})$. $C_i$ is given for the SOC objective, by
% \begin{equation}
% \label{eq:meeting_cost_mksp}
%     C_i^{\text{MKSP}}(v,t) = \left\{\begin{array}{cc}
%          t + d\argument{v, g_i}, & t \geq \earliestvertextime{i}\\
%          \infty, & \text{otherwise}
%     \end{array}\right.,
% \end{equation}
%and
\begin{equation}
\label{eq:meeting_cost_soc}
    C_i(v,t) = \left\{\begin{array}{cc}
         2\cdot t + d\argument{v, g_i}, & t \geq \earliestvertextime{i}\\
         \infty, & \text{otherwise}
    \end{array}\right.,
\end{equation}
where $\earliestvertextime{i}$ is the earliest possible meeting time at $v$ for task $\task{i}$, i.e., the earliest time both assigned agents can arrive at $v$. Specifically, $\earliestvertextime{i}$ is defined as
\begin{equation}
\label{eq:earliest_meeting_time}
\begin{split}
    \earliestvertextime{i} = \max\left\{
    d\argument{\agentstartfunc{\agentAi{i}}, s_i} + d\argument{s_i, v}, \right.\\
    \left.d\argument{\agentstartfunc{\agentBi{i}}, v}\right\}
\end{split},
\end{equation}
where~$d(u,v)$ is the length of the single-agent shortest path from $u$ to $v$. 
If~${d(u,v)=\infty}$, no such path exists.

The first step of \ouralg is to compute~$\meetingtable{i}{}$, the meetings table for each task~$\task{i}$ (lines 3-4). 
The meetings table is a function~${\meetingtable{i}{}: V \rightarrow \mathbb{R}\cup\set{\infty}}$ that returns for each vertex~${v\in V}$ the cost for completing task~$\task{i}$ with a meeting in~$v$ at the earliest possible time.
${\meetingtable{i}{}\argument{v}}$ is initialized for each~${v \in V}$ with~${\meetingtable{i}{}\argument{v} = C_i(v,\earliestvertextime{i})}$. %and~${\meetingtable{i}{\text{SOC}}\argument{v} = C_i^{\text{SOC}}(v,\earliestvertextime{i})}$.
Each meetings table is stored as a heap which allows for $insert$, $update$ and $getMin$ operations in $\calO(\log\setsize{V})$.
These operations are used during the root node expansion which will be described shortly.

We compute $\meetingtable{i}{}\argument{v}$ for all $v\in V$ in polynomial time using \astar and \dij's algorithm as described in Algorithm~\ref{alg:meetings_table}.
Computing the meetings table for each task $\task{i}$ requires finding paths from every node $v\in V$ to the agents' start locations, as well as tasks' start and goal locations. 
Given the meeting tables, we can check if the given problem instance is well-formed (see Definition \ref{def:well-formed}). 
More specifically, it is well-formed iff for every meetings table $\meetingtable{i}{}$ there exists a vertex $v \in V \setminus V_{ep}$ such that $\meetingtable{i}{}\argument{v}<\infty$. 

\subsubsection{Root initialization}
We define the cost of a set of meetings~${\meetingsetset}$ as follows:~%${C^{\text{MKSP}}\argument{\meetingset} = \max_{1\leq i \leq \numagents}{C_i^{\text{MKSP}}\meeting{i}}}$ and~
${C\argument{\meetingset} = \sum_{i=1}^{\numagents}{C_i\meeting{i}}}$.
$\meetingset^*$ is an optimal set of meetings that minimizes the problem objective while ignoring possible conflicts between agents.
Namely, 
\begin{equation}
\label{eq:optimal_meetings_set}
\meetingset^*\in \argmin_{\meetingset}{C\argument{\meetingset}}.
\end{equation}

\ouralg's search starts with creating the initial CT root node with an empty set of constraints, and an optimal set of meetings~${\meetingset^*}$, by choosing a lowest-cost meeting for each task from the meeting tables (lines 5-7). 
Given $\meetingset^*$, the paths level is called to compute individual paths for each agent (line~8). 
This is similar to \cbs, except that in the path-level search we plan for each task~$\task{i}$ in parts: (i)~for~${\agentAi{i}}$ from~${\agentstartfunc{\agentAi{i}}}$ to~$s_i$, and then from~$s_i$ to~${\meetinglocation{i}}$ at time $\meetingtime{i}$, and (ii)~for~${\agentBi{i}}$ from~${\agentstartfunc{\agentBi{i}}}$ to~${\meetinglocation{i}}$ at time~${\meetingtime{i}}$ and then to~$g_i$.\footnote{For simplicity, we assume a disappear-at-target behavior \cite{stern2019}, such that the {\agentAname} agent disappears after the meeting, and the {\agentBname} agent disappears after completing the task (at the task goal location).}
%Note that when planning for the meeting, the goal is not just the meeting location but also the meeting time.
Note that when planning for a meeting, we should consider both the meeting location and time.
% Note that planning for the meeting has a goal location and time.
The initial CT root node cost is computed and it is inserted to the \textsc{Open} list (lines~{9-10}).

\subsubsection{Selecting a node for expansion}
As long as there are nodes in the \textsc{Open} list (line~11), we follow \cbs's best-first search approach and select a node with a lowest cost (line~12). 
If the \textsc{Open} list contains both root and regular nodes with the same lowest cost, \ouralg chooses to expand a regular node (to perform this in practice, \ouralg keeps root and regular nodes in two separate \textsc{Open} lists).
% This way, \ouralg avoids creating a large number of CT root nodes before trying to find feasible paths for a given 
% \ouralg uses a tie-breaking mechanism:
% it randomly chooses (with a pre-defined probability $p_{root}$ to select a root node) the type of node to select, i.e., root or regular (to perform this in practice, \ouralg keeps root and regular nodes in two separate \textsc{Open} lists). 
% From our experience, this tie-breaking mechanism provides a good exploration-exploitation trade-off of the meetings space: the algorithm is given a chance to either exploit a given set of meetings to try find and collision-free paths, or explore the meetings space by checking the next best set of meetings.

\subsubsection{Expanding a root node}
\label{sec:expand_root}
After selecting a lowest-cost node~$N$ from the \textsc{Open} list, \ouralg checks for conflicts in its solution (line 13). 
If none are found, $N.solution$ is returned as the optimal solution (lines 14-15). 
Otherwise, if~$N$ is a root node, it is expanded to get its successors in the meetings space (lines 16-17). 
The process of expanding a root node is described in Algorithm~\ref{alg:expand_root}. 
% Generally, it creates~$k$ new root nodes (as the number of tasks), each of them is created by replacing the meeting of a single task with the next best meeting available for the assigned pair of agents.
Given the current set of meetings (in the expanded root node)~${\meetingsetset}$, \ouralg generates up to $\numagents$ new sets of meetings, one for each task.
This is done in a non-decreasing manner, by replacing one meeting $m_i\in\meetingset$ at a time, an idea similar to the Increasing Cost Tree Search (\icts) \cite{sharon2013} algorithm, thus creating $\numagents$ new root nodes.

% Getting the next best meeting (for each task at a time), is how \ouralg performs the meetings-level search in the meetings space.
% The meetings space has a time dimension, i.e., it is an~$(n+1)-$~dimensional space, where $n$ is the number of dimensions in the paths space.
% For example, for a 2-D grid environment, the meetings space is in 3-D (i.e.,~${x, y}$ and~$t$).
To get the next-best meeting for task $\task{i}$, we have to search both for different locations and time steps in the meetings space.
The meetings table $\meetingtable{i}{}$ of $\task{i}$ initially consists of meetings at each possible location, at the earliest time possible. %, and sorted by meeting cost.
Each time \ouralg invokes the get-next-meeting procedure for $\task{i}$ (line 6 in Algorithm \ref{alg:expand_root}), it returns the lowest-cost meeting~${m_i=\meeting{i}}$ from $\meetingtable{i}{}$.
The table is then updated so that it holds the next lowest-cost meeting.
This is done by updating~${\meetingtable{i}{}\argument{\meetinglocation{i}} =C_i\argument{\meetinglocation{i},\meetingtime{i}+1}}$. 
Namely, updating the cost of meeting at~${\meetinglocation{i}}$, but at time~${\meetingtime{i}+1}$ rather than~${\meetingtime{i}}$.
% The meetings table is then re-sorted so that the next best meeting is returned the next time the get-next-meeting procedure is invoked.
The next time the get-next-meeting procedure is invoked, the next best meeting will be returned by the table.

Subsequently, a new path is planned for the pair of agents whose meeting changed, the new CT node cost is computed and it is inserted into the \textsc{Open} list.

\subsubsection{Resolving a conflict}
The last part of the algorithm is almost identical to \cbs: when expanding a node $N$ (either root or regular) \ouralg splits its CT and creates a regular node for each agent by the first conflict found (lines 18-19). 
These nodes has the same set of meetings as $N$ (line 22).

% %%%%%%%%%%%%%%%%%%%%%%%%%%%%%%%%%%%%%%%%%%%%%%%%%%%%%%%%%%%%%%%%%%%%%%%%%%%%%%%%
\section{Theoretical Analysis}
\label{sec:theory}
\subsection{\ouralg Completeness}
We restrict our discussion to well-formed {\cmapf} instances. 
We guarantee completeness of \ouralg on well-formed instances using Claim~\ref{claim:well-formed}, which states that if the instance is not well-formed we can identify it by running a polynomial-time test procedure before executing \ouralg.

% \footnotetext{The result of this computation is included in line 4 as well, but if~${\agentstartfunc{\agentAi{i}}}$ and~$s_i$ are not connected, then the problem is not well-formed and we can stop the algorithm to save time.}

\begin{theorem} \label{theorem:completness}
\ouralg will return a solution for any well-formed {\cmapf} instance.
\end{theorem}

\begin{proofs}
By Lemma~\ref{lemma:well-formed} we know that there exists a solution. 
Denote the set of meetings which forms the solution by~${\meetingset=\set{\meeting{1}, \meeting{2},\dots,\meeting{k}}}$. 
\ouralg's meetings level preforms a systematic best-first search across the meetings space, thus, will eventually create a root node, denoted by~$R_\meetingset$, whose set of meetings is~$\meetingset$. 
There exists a feasible solution such that each pair of agents~${\argument{\agentAi{i},\agentBi{i}}}$ meet at~${\meeting{i}}$.
By the completeness of \cbs it is guaranteed that the search from the CT root node~${R_\meetingset}$ will eventually find the solution.
\end{proofs}

% \begin{figure*}[ht]
%     \centering
%     \begin{subfigure}{0.5\columnwidth}
%         \includegraphics[width=\textwidth]{figures/results/Success_Rate,_empty-48-48_SOC.png}
%         \caption{empty}
%     \end{subfigure}
%     \begin{subfigure}{0.5\columnwidth}
%         \includegraphics[width=\textwidth]{figures/results/Success_Rate,_den312d.png}
%         \caption{DAO}
%     \end{subfigure}
%     \begin{subfigure}{0.5\columnwidth}
%         \includegraphics[width=\textwidth]{figures/results/Success_Rate,_maze-32-32-4.png}
%         \caption{maze}
%     \end{subfigure}
%     \begin{subfigure}{0.5\columnwidth}
%         \includegraphics[width=\textwidth]{figures/results/Success_Rate,_warehouse-10-20-10-2-1.png}
%         \caption{large warehouse}
%     \end{subfigure}
%     % \begin{subfigure}{0.19\textwidth}
%     %     \includegraphics[width=\textwidth]{figures/results/Success_Rate,_small-warehouse-26-15.png}
%     %     \caption{small warehouse}
%     % \end{subfigure}
%     \caption{Success rate with different $p_{root}$ values.}
%     \label{fig:p_root_sucess_rate}
%     \vspace{-0.4em}
% \end{figure*}

\subsection{\ouralg Optimality}
We again restrict the discussion to well-formed instances. 
By Theorem~\ref{theorem:completness} we are guaranteed that \ouralg solves every well-formed {\cmapf} instance.
We show that it returns an optimal solution for the SOC objective function.
% A similar proof can be applied for SOC as well.

\begin{lemma} \label{lemma:alg-non-decreasing}
Let $\meetingset$ be a set of meetings with~${C\argument{\meetingset}=c}$ and let~$N$ be a CT node with cost larger than~$c$.
\ouralg will generate a root node corresponding to~$\meetingset$ before expanding~$N$.
\end{lemma}

\begin{proofs}
Assume that there exists a set of meetings~$\meetingset$ s.t.~${C(\meetingset) = c}$, that hasn't been generated yet.
Assume by contradiction that \ouralg expands a node~$N$ that has a solution with cost~${c' > c}$.
By definition, the first set of meetings $\meetingset_0$ that is generated (line 6 in Algorithm~\ref{alg:co-cbs}) induces a solution which minimizes the SOC objective function. 
In particular, this implies that the cost of completing all tasks in the (possibly infeasible) solution induced by~${\meetingset_0}$ is less than or equal to~$c$. 
Similarly, the cost of completing all tasks in the solution induced by~${\meetingset}$ is less than or equal to~$c$. 
Therefore, there exists a sequence of meeting placements that may be generated during the meetings-level search, from~${\meetingset_0}$ to~${\meetingset}$, such that the cost of the meetings sets in the sequence would remain smaller or equal to $c$ at all time, i.e., each set of meetings $\meetingset'$ in this sequence holds~${C(\meetingset') \leq c}$. 
The way the meetings-level search works ensures that there must be at least one root node in the \textsc{Open} list consisting of one of these meeting sets. 
Therefore, there exists a root node that hasn't been expanded yet in the \textsc{Open} list with a cost smaller than~$c'$, in contradiction to the best-first search approach which chose node~$N$ with a larger cost for expansion.
\end{proofs}

% \begin{figure}[ht]
%     \centering
%     \begin{subfigure}{0.49\columnwidth}
%         \includegraphics[width=\textwidth]{figures/results/Success_Rate_MKSP.png}
%         \caption{}
%         \label{fig:success_rate_mksp}
%     \end{subfigure}
%     \begin{subfigure}{0.49\columnwidth}
%         \includegraphics[width=\textwidth]{figures/results/Success_Rate_SOC.png}
%         \caption{}
%         \label{fig:success_rate_soc}
%     \end{subfigure}
%     % \setlength{\abovecaptionskip}{2pt}
%     \caption{Success rates for (\subref{fig:success_rate_mksp}) MKSP and (\subref{fig:success_rate_soc}) SOC.}
%     % \caption{Success rates for MKSP (left) and SOC (right).}
%     \label{fig:success_rate}
%     \vspace{-0.4em}
% \end{figure}

\begin{theorem}
\ouralg returns an optimal solution for any well-formed {\cmapf} instance.
\end{theorem}

\begin{proofs}
Assume that there exists an optimal solution with some cost $c^*$. \ouralg performs a \cbs-like search on each generated CT, namely, it searches through a forest of conflict trees. 
By Lemma~\ref{lemma:alg-non-decreasing} we get that the cost of each expanded root node of each CT constitutes a lower-bound on~$c^*$. 
From the optimality guarantees of \cbs, we get that any node expanded in each of those CTs (i.e., regular nodes) is also a lower bound on~$c^*$.
Due to \ouralg's best-first approach, it won't expand a node with a cost larger than~$c^*$ before completing a search through all possible CT nodes with cost~$c^*$ (by expanding neither a root node nor a regular one). 
Since there exists a solution with such cost, and the number of possible solutions with a specific cost is finite, \ouralg will eventually expand a node with an optimal and feasible solution and return it. 
\end{proofs}

\subsection{\ouralg Runtime Analysis}
% \ouralg extends \cbs by adding a level that searches in the meeting space. 
\ouralg is an extension of \cbs which adds a level that searches in the meeting space. 
The size of the meeting space is~${\calO((|V| \cdot c^*)^k)}$. 
In the worst case, \ouralg would generate all possible meetings and perform a \cbs search for each set of meetings (up-to cost $c^*$). 
It may result in a number of expanded nodes relative to the number of meeting points, times the number of nodes expanded by \cbs~{\cite{sharon2015,gordon2021}}.

In practice, in our empirical evaluation (Section \ref{sec:experiments}) we observe that the number of generated meeting sets is typically small, especially for large or sparse environments (see Fig.~\ref{fig:generated_meetings}).
This means that the number of full \cbs searches (one for each meetings set) is usually small.
However, in scenarios with many conflicts, a large number of root nodes (and, meeting sets) are created.
This causes an increase in run time which is exponential in the number of tasks.
%can choose a node for expansion of either types--a root node which will result in inserting new root nodes (with new meetings) to the \textsc{Open} list, or a regular node (by splitting the CT node according to a conflict). 
%Thus, \ouralg may expand all possible root nodes with a cost up-to $c^*$ (the optimal solution's cost), and for each such root node it'll perform a full \cbs search up-to the same cost. 
%We emphasize that in the general case it is reasonable to expect several sets of meetings which will imply a feasible solution with an optimal cost. 
%Therefore, if we'll allow the algorithm to combine between expanding root nodes and regular nodes (while preserving the monotonically non-decreasing expansion order), it'll usually find a solution much faster. 
%For that, we use the aforementioned tie-breaking mechanism, in which the algorithm chooses uniformly if to expand a root node or a regular node. 
%This mechanism creates a good balance between creating new CTs and searching existing CTs, preventing it from getting ``stuck'' in one direction of the search.

% \subsection{Other solution approaches}
% \label{sec:comparison}
% \ask{TODO: move to evaluation}

% %%%%%%%%%%%%%%%%%%%%%%%%%%%%%%%%%%%%%%%%%%%%%%%%%%%%%%%%%%%%%%%%%%%%%%%%%%%%%%%%

\section{\ouralg Improvements}
\newtext{
\label{sec:improvements}
In the previous section, we introduced the basic version of \ouralg for solving the {\cmapf} problem.
\ouralg creates a forest of conflict trees and runs \cbs on each tree.
Thus, we can apply previously-suggested \cbs improvements to \ouralg.
% These improvements and optimization techniques have shown the potential to decrease the \cbs's run-time, and are 
One such improvement that has been shown to significantly decrease \cbs's run-time is \emph{prioritizing conflicts (PC)} \cite{boyarski2015}.
In this section we present in details the application of PC to \ouralg.
More \cbs improvements are discussed in Section~\ref{sec:discussion}.
In addition, we introduce a unique improvement for \ouralg called \emph{Lazy Expansion (LE)}, which exploits special characteristics of root nodes.
Both improvements keep \ouralg optimal, while introducing a significant improvement in run time, as shown empirically in Section~\ref{sec:experiments}.
}

\subsection{Prioritizing Conflicts (PC) for \ouralg}\
\newtext{
\label{sec:PC}
The Improved \cbs (\icbs) algorithm \cite{boyarski2015} introduced an enhancement to \cbs by defining rules dictating how to split the CT.
In particular, conflicts are divided into three types: \emph{cardinal}, \emph{semi-cardinal} and \emph{non-cardinal}.
Cardinal conflicts always cause an increase in the solution cost, therefore \icbs chooses to split cardinal conflicts first.
Cardinal conflicts are identified by examining the width of a \emph{multi-value decision diagram (MDD)} \cite{sharon2013}, which is constructed for each low-level path found.
The MDD is a directed a-cyclic graph which compactly stores all possible paths of a given cost~$c$ for a given agent, from its start vertex to its goal vertex.
An MDD of cost~$c$ consists of~$c$ layers, corresponding to~$c$ time steps.%while completely ignoring the other agents.
}

\newtext{
Applying PC to \ouralg is not straightforward, since an MDD stores paths from a start vertex to a goal vertex, while in {\cmapf} paths are constrained to ensure cooperation between agents.
% More specifically, in our {\cmapf} setting, each valid path has two phases.
% A path either traverses the task start location on its way to the meeting (for the {\agentAname}), or it traverses the meeting location (at a specific time) on its way to the goal (for the {\agentBname}).
More specifically, in our {\cmapf} setting, each agent has an \emph{intermediate goal}, i.e., the task start location, or the meeting location (at a specific time).
% % We need to take this into consideration while building the MDD.
We therefore need to modify the way an MDD is constructed, and indeed we suggest a method for efficiently doing so for both agents.
}

\newtext{
For the {\agentAname} agent, we must ensure it passes through the task's start location.
% that all paths traverse via the task start location.
In other words, we need to prune MDD nodes that are not part of any of the agent's paths which pass the task's start location. 
We refer to such nodes as \emph{invalid nodes}.
%that traversing them does enable traversing the task start location while arriving at the meeting.
% The way for constructing the MDD graph efficiently,
Constructing an MDD efficiently is done using two breadth-first searches--one forward and one backward (start to goal and vise versa)~\cite{sharon2013}.
%To efficiently construct an MDD given a start, goal and path length, we perform two breadth-first searches--one forward and one backward (start to goal and vise versa)~\cite{sharon2013}.
% In order to efficiently prune invalid nodes, we mark the MDD nodes with two flags--\emph{valid\_forward} and \emph{valid\_backward} during the forward and backward passes, respectively.
% We mark a node as valid (forward or backward) if it can be reached from a node representing the task's start location.
In order to efficiently prune invalid nodes, we follow the following procedure: during the forward search, we mark MDD nodes corresponding to the task start location and all their descendants as \emph{valid\_forward}.
Similarly, during the backward search, we mark these nodes and all their ancestors as \emph{valid\_backward}.
% More specifically, for a task~$\task{i}$ with a start location~$s_i$, during the forward pass we mark every MDD node~$v$ with location~$s_i$ as \emph{valid\_forward}.
% Each decedent of~$v$ will also be marked as \emph{valid\_forward}, and so on.
% A similar process is done during the backward pass, by marking~$v$ as \emph{valid\_backward}, and all its ancestors.
Finally, all MDD nodes that are not marked with either flags are pruned.
}
% \ask{TODO: (above) this is a lot of text that is a bit hard to follow. Maybe simplify via notation. Consider a specific task and denote the task's start location by some letter> I think that will help}

\newtext{
For the {\agentBname} agent, constructing the MDD requires only slight changes.
We need to constrain the agent to be at the meeting's location at the meeting's time.
We simply do it by eliminating all other nodes from the MDD layer corresponds to the meeting time during the forward pass in the MDD construction. 
}
% \OG{I feel that there is not enough explanation about the MDDs, but maybe it is just because I'm super-involved with them in my work. I think that we at least need to mention the "layers" of the graph, and say that they are corresponding to timesteps}

\subsection{Lazy Expansion (LE) of root nodes}\
\newtext{
\label{sec:LE}
\ouralg searches the meetings space by creating root nodes, each corresponding to a unique set of meetings.
Note that since no constraints are imposed on paths of root nodes, their cost is given as an aggregation of their meeting costs.
Furthermore, meeting costs are computed a-priori during the construction of meeting tables (see Section~\ref{sec:algorithm}).
% \ouralg searches the meeting space via the notion of root nodes (see Section~\ref{sec:algorithm}).
% Each root node has a unique set of meetings, one meeting for each task.
% We observe two special characteristics of root nodes: (i)~being the root of a CT, root nodes have no constraints imposed on low-level paths 
% \OG{not clear, I'm not sure I understood what you trying to say in 1}
% , and (ii)~due to that, their solution cost is known a priori as an aggregation of the meeting costs, previously computed during the construction of meeting table (see Section~\ref{sec:meeting_tables}.
% \OG{can't we simply say "observe that the cost of a root node is known a priori from the calculation of the meetings set (and maybe refer to the part in the algorithm)?}
This means that when a root node is expanded, and new root nodes are created, they can immediately be inserted into the \textsc{Open} list \emph{without} computing their low-level paths.
The low-level paths will be computed only when these root nodes are extracted from the \textsc{Open} list. We term this \emph{Lazy Expansion (LE)}.}
% \ask{TODO: rephrase the above paragraph}

\newtext{
Each time a root node is expanded, it creates $\numagents$ new root nodes by replacing the meeting of each of the tasks.~We emphasize that while generating those nodes is mandatory in order to guarantee optimality, most of them won't be expanded. Thus, the run-time saved by LE can be significant.
}
\vspace{-0.3em}
% %%%%%%%%%%%%%%%%%%%%%%%%%%%%%%%%%%%%%%%%%%%%%%%%%%%%%%%%%%%%%%%%%%%%%%%%%%%%%%%%
\section{Experimental Evaluation}
\label{sec:experiments}

\newtext{
\ouralg solves the newly introduced {\cmapf} problem.
To the best of our knowledge, there does not exist an off-the-shelf optimal solver for MAPF problems involving cooperative behavior.
% Therefore, it is not possible to measure the performance of \ouralg against an off-the-shelf algorithm.
Suggesting a centralized \astar-based implementation for solving the {\cmapf} problem is challenging due to constraints imposed on low-level paths to achieve cooperation.
Such approach would require to perform a search in the meetings space, resulting in an exponentially-large state space.
Moreover, an attempt to solve {\cmapf} using such implementation would yield similar results as solving classical MAPF problem using \astar \cite{sharon2015}, due to their similar search approach and conflict-resolution mechanism.
Thus, we restrict our empirical evaluation to the algorithms presented in this paper.
}

\changed{To measure the quality of \ouralg, we present the results of an empirical evaluation performed on standard MAPF benchmarks \cite{stern2019,sturtevant2012} showing the performance of the basic version of \ouralg, as well as the two suggested improvements (see Section~\ref{sec:improvements}).
% As stated in Section~\ref{sec:comparison}, it is not obvious 
% \OG{I think that saying that weakens the results, maybe think of a different phrasing} 
% how to measure the quality of Co-CBS because there is no possible comparison with previous approach.
% Since {\cmapf} introduces a new problem, it is not possible to measure the performance of \ouralg against a previous approach.
% Furthermore, even a centralized \astar-based approach is not straightforward, due to the cooperation constraints that need to be imposed on solutions.
% Sub-optimal methods (e.g., a three-phase \astar search to a fixed set of meetings) are computationally intractable and perform poorly on trivial problem instances.
% We provide an empirical evaluation of \ouralg on standard MAPF benchmarks \cite{stern2019,sturtevant2012}, showing the performance of the basic version of \ouralg, as well as the two suggested improvements (see Section~\ref{sec:improvements}).
\ouralg is implemented in C++\footnote{Upon acceptance, we will make the code publicly available.} and is based on the implementation of Li et al.~\cite{li2021}. %Python\footnote{https://github.com/CRL-Technion/Cooperative-MAPF}. 
All simulations were performed on an Intel Xeon Platinum~{8000}~@~{3.1}Ghz machine with~{32.0 GB RAM}.
}

% In this section we present the results of an empirical evaluation of \ouralg on standard MAPF benchmarks \cite{stern2019,sturtevant2012}.
% \ouralg is implemented in Python\footnote{https://github.com/CRL-Technion/Cooperative-MAPF}. All simulations were performed on an Intel Xeon Platinum 8000 @ 3.1Ghz machine with 64.0 GB RAM.

\subsection{Benchmarks and setup}

\changed{
We evaluated \ouralg on several 2D grid-based benchmarks. Specifically, we tested \ouralg on different types of maps---a dense game map (\textit{DAO, den312d}), random map (\textit{random-32-32-20}), a large warehouse {(\textit{warehouse-10-20-10-2-1})} and a custom small warehouse ($57\times 27$).
% These benchmarks allow us to estimate the performance of \ouralg on realistic environments where cooperation is needed. \NGR{think about this sentence.}
We ran~25 random queries for each benchmark for the SOC objective with the number of tasks ranging from $6$ tasks ($12$ agents) to~$22$ tasks (44 agents) and with a timeout of two minutes.
%In order to evaluate how the number of tasks affects the algorithm's performance, we ran each such scenario for a different number of tasks, on a scale from 1 (two agents) to 8 (16 total agents).
On each benchmark, we compare the performance of three different variances of \ouralg: (i)~basic \ouralg, (ii)~\ouralg with prioritizing conflicts (PC), and (iii)~\ouralg with PC and lazy expansion (LE) of root nodes.
}
As opposed to the classical MAPF, where each agent is provided with start and goal locations, in {\cmapf}, a task's start and goal need to be provided (instead of explicitly providing an agents' goal).
Thus, we defined the tasks in each scenario as follows, based on the original benchmark scenario: 
for each pair of agents, one set of start and goal locations is used for the task, and the other set is used for the agents' start locations.

% maze-32-32-4 (790)
% small-warehouse20-10 (140)
% Note that in classical MAPF, a low-level solution is found using $\numagents$ \astar searches (one per agent).
% In {\cmapf}, however, we plan two paths for each agent (the {\agentAname} to the task start location and then to the meeting location, the {\agentBname} to the meeting location and then to task goal location).
% Therefore, finding a {\cmapf} solution for $\numagents$ tasks is equivalent to $4\numagents$ agents in classical MAPF, in terms of time complexity.
% \NGR{runtime of a computing the solution of a single node}

\begin{figure*}[ht]
    \centering
    \begin{subfigure}{0.245\textwidth}
        \includegraphics[width=\textwidth]{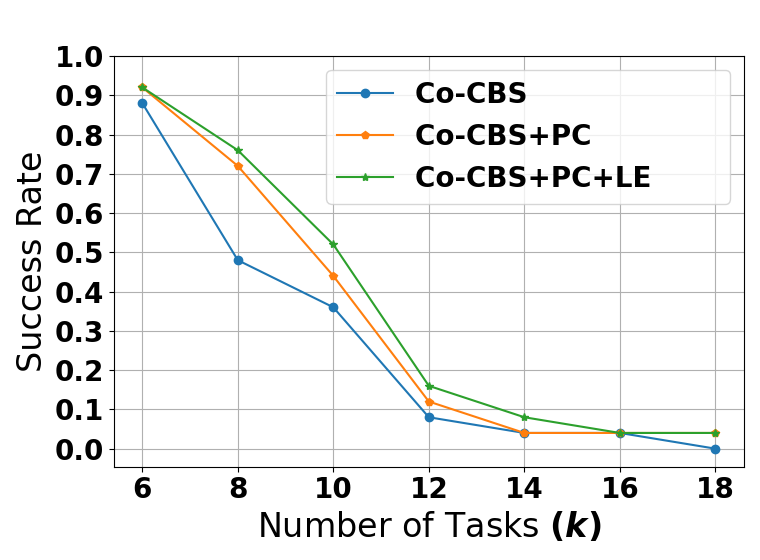}
        \caption{den312d}
        \label{fig:den312d_Success_Rate}
    \end{subfigure}
    \begin{subfigure}{0.245\textwidth}
        \includegraphics[width=\textwidth]{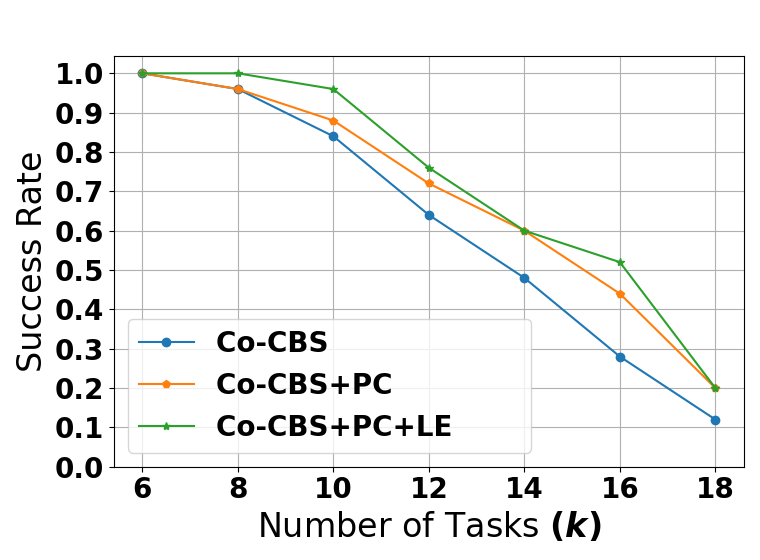}
        \caption{random-32-32-20}
        \label{fig:random-32-32-20_Success_Rate}
    \end{subfigure}
    \begin{subfigure}{0.245\textwidth}
        \includegraphics[width=\textwidth]{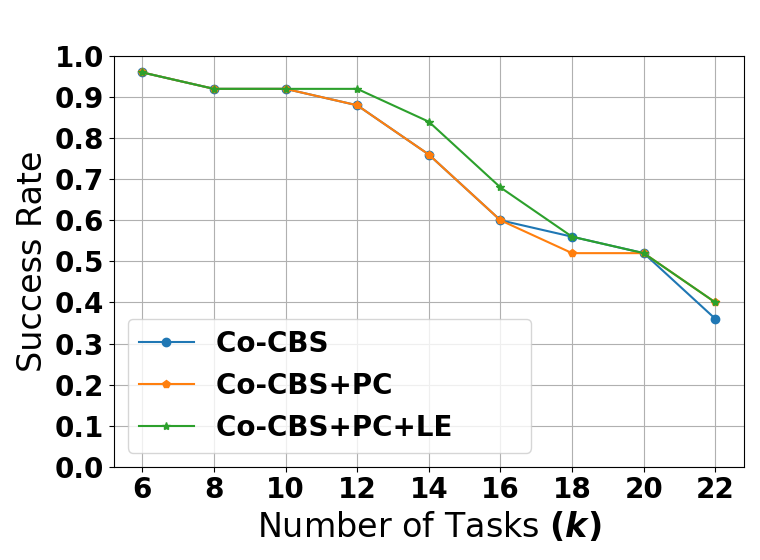}
        \caption{warehouse-10-20-10-2-1}
        \label{fig:warehouse-10-20-10-2-1_Success_Rate}
    \end{subfigure}
    \begin{subfigure}{0.245\textwidth}
        \includegraphics[width=\textwidth]{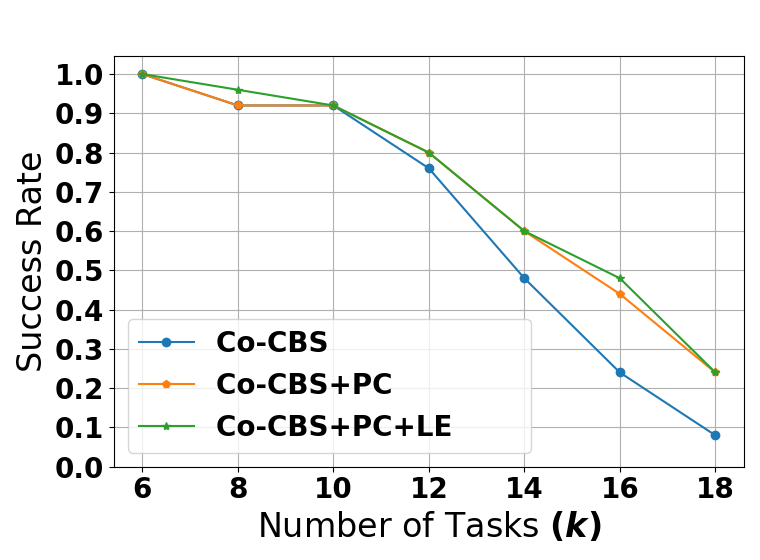}
        \caption{warehouse-57-27}
        \label{fig:warehouse-57-27_Success_Rate}
    \end{subfigure}
    \caption{Success rates.}
    % \caption{Success rates for MKSP (left) and SOC (right).}
    \label{fig:success_rate_dense}
    \vspace{-0.4em}
\end{figure*}

\begin{figure}[ht]
    \centering
    \begin{subfigure}{0.49\columnwidth}
        \includegraphics[width=\textwidth]{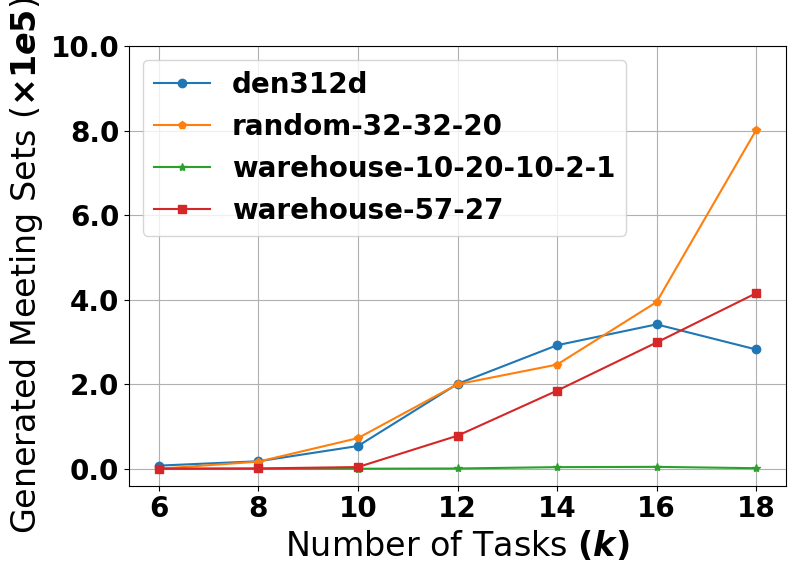}
        \caption{}
        \label{fig:generated_meetings}
    \end{subfigure}
    \begin{subfigure}{0.49\columnwidth}
        \includegraphics[width=\textwidth]{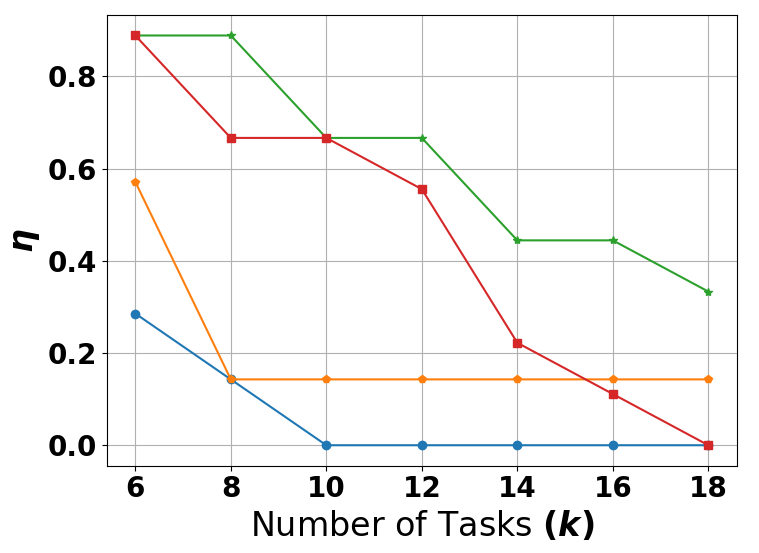}
        \caption{}
        \label{fig:first_meetings}
    \end{subfigure}
    \setlength{\abovecaptionskip}{2pt}
    \caption{(\subref{fig:generated_meetings}) Number of generated sets of meetings. (\subref{fig:first_meetings}) Ratio~$\eta$ between the number of instances solved using the first set of meetings, and the total number of instances.}
    % \caption{Ratio between solutions with the first set of meetings and all solutions, denoted $\eta$ (left). Number of generated sets of meetings (right).}
    \label{fig:first_generated_meetings}
    % \vspace{-2.0mm}
\end{figure}

\subsection{Results}
% We ran $25$ random scenarios for each map with the MKSP objective. \OG{If you choose to replace the previous subsection with my suggestion, then start this one from the next line (and add the Makespan to the previous section)}
\changed{
We first examine the algorithm's success rate (i.e., the ratio of solved instances within the time limit) for all benchmarks.
% Figures~\ref{fig:success_rate_mksp} and~\ref{fig:success_rate_soc}
Figures~\ref{fig:success_rate_dense} shows the success rates of \ouralg on all maps.
\ouralg successfully solves more than~$80\%$ of the instances (excluding the den312d benchmark) with ten tasks.
The success rate sharply drops below~$20\%$ for twelve tasks or more on the dense den312d map.
}
\newtext{
Using PC improves the basic \ouralg in all cases, achieving up to $30\%$ increase in the success rate.
Furthermore, adding LE on top of PC further improves the performance in most cases, and never degrades the performance.
This is especially notable with a large number of tasks, where many root nodes are created.
}
\changed{
% We now continue to examine the number and types of meeting sets generated.
Fig.~\ref{fig:generated_meetings} shows the average number of generated meeting sets.
Fig. \ref{fig:first_meetings} shows the ratio~$\eta$ between the number of instances where the first set of meetings is used to obtain the solution and the total number of instances.
Both warehouse environments are typically sparser, causing fewer conflicts between agents.
Thus, a feasible solution is usually quickly found using the first set of meetings.
% since there are typically very few conflicts between agents.
This is especially true in the large warehouse, 
% where most solutions are obtained using the first set of meetings 
namely when $\eta$ is close to one.
% , as seen in Fig.~\ref{fig:first_meetings}).
The search in this case is equivalent to running \cbs with the first set of meetings.
For the same reason, PC does not improve the performance in this environment.
% over the basic \ouralg.
% This is also the reason why it is the only environment in which PC does not achieve an improvement over the basic \ouralg.
Applying LE as well, however, does manage to improve the success rate for the majority of tasks.
In other environments, on the other hand, maps are smaller and denser, and most solutions are not obtained using the first generated set of meetings.
A more exhaustive meeting-space search is therefore required to find an optimal solution, as shown in Fig.~\ref{fig:generated_meetings}.
}

% Finally, figure \ref{fig:num_expansions} shows the number of expanded nodes by type (regular or root) on a logarithmic scale.
% The large number of expanded root nodes (represented by the dark bars) expresses the exhaustive meeting-space search.
%\vspace{-0.4em}

% Fig. \ref{fig:generated_meetings} corresponds with this, showing that the number of generated meeting sets grows for denser maps and large number of tasks.

% number of expanded nodes and the number of generated meeting sets, respectively, for~${p_{root}=0.2}$.
% We see that in sparse environments \ouralg converges to a solution after a relatively small number of iterations while generating only a small number of meeting sets.
% In such environments there exists large number of possible meeting sets that can result in an optimal solution.
% A feasible solution is therefore quickly found since there are few conflicts between agents.   
% In large maps, an optimal solution usually consists of a small number of conflicts.
% This is emphasized in the empty map where the success rate is very high.
% Thus, the search focuses on a small number of meeting sets and performs a \cbs search with each set.
% On a smaller and denser map (i.e., the small warehouse), on the other hand, there are a lot of conflicts between agents.
% Thus, \ouralg expands a large number of meeting sets before finding a feasible solution.
% We also observe that the number of generated meeting sets increases with the number of agents (as expected).

% %%%%%%%%%%%%%%%%%%%%%%%%%%%%%%%%%%%%%%%%%%%%%%%%%%%%%%%%%%%%%%%%%%%%%%%%%%%%%%%%
\section{Discussion and Future Work}
\label{sec:discussion}

In this paper, we introduced the Cooperative Multi-Agent Path Finding ({\cmapf}) problem, an extension to classical MAPF that incorporates cooperative behavior. 
We introduced \ouralg, a three-level search algorithm that optimally solves {\cmapf} instances, \newtext{as well as two improvements, Prioritizing Conflict (PC) and Lazy Expansion (LE).}
%for cases where the cooperation between the agents is defined by a set of cooperative tasks.  Each task requires a pre-assigned pair of agents to meet during execution.

In this section, we provide a comprehensive discussion regarding the suggested model and algorithm.
Specifically, we discuss further possible improvements that can be applied to \ouralg and suggest possible extensions to the {\cmapf} model.
We argue that \ouralg forms a basic framework that may serve as a starting point for future extensions.
% \vspace{-0.5em}

\subsection{\ouralg's Extensions and Improvements}
% \subsubsection{Node selection and expansion}
% \ouralg may create a very large number of root nodes, as the meetings space is exponential in the size of the graph and number of tasks.
% It may be beneficial to use a heuristic function to guide the search towards better meeting sets, as well as a pruning mechanism to reject possibly-infeasible or sub-optimal meeting sets.
% %\vspace{-0.2em}

\subsubsection{Information reusing between conflict trees}
\ouralg expands root nodes by only changing one meeting in the newly-created node.
Moreover, the next selected meeting is usually very close to the current meeting, both in location and time.
This implies that \ouralg searches over multiple trees that potentially have very similar solutions.
We may exploit this for more efficient computation.
%\vspace{-0.2em}

\subsubsection{Meetings-level search}
\ouralg uses a simple-yet-effective method for finding an optimal meeting for each task.
For large problem instances, this method may become memory and run-time expensive, due to the maintenance of large meeting tables.
We may consider incorporating an algorithm such as the recently-proposed \mmstar algorithm \cite{atzmon2020b}, for the Multi-Agent Meeting problem.
Furthermore, we may couple meetings and paths planning, and handle conflicts during the search for a meeting.
This may be advantageous as meetings and conflicts may be tightly coupled.

% \subsubsection{Two-level planning}
% \ouralg decouples meetings planning from path planning by adding a third level of planning.
% Using \mmstar, as suggested, to simultaneously search for a meeting for all agents can be used to handle conflicts between agents during the search for a meeting. Coupling the two may be advantageous as meetings and conflicts may be tightly coupled.
% Planning can then be done in two levels, where the low level handles both meetings and constraints.

%\subsubsection{Parallel \ouralg}
%\ouralg's search makes it a good candidate for parallelization.
%We may perform a parallel search on multiple conflict trees, which is equivalent to running multiple \cbs searches simultaneously, each with a given set of meetings. 
% \newtext{
% \subsubsection{Concurrent and Distributed Planning}
% \ouralg's search makes it a good candidate for parallelization.
% We may perform a parallel search on multiple conflict trees, which is equivalent to running multiple \cbs searches simultaneously, each with a given set of meetings.
% Furthermore, distributed approaches may be considered for solving (sub-optimally) the {\cmapf} problem.
% Such an approach is presented by Bucchiarone, \cite{bucchiarone19}, where plans are computed concurrently for multiple heterogeneous agents, and agents form \emph{ensembles} to resolve conflicts. 
% }

\subsubsection{Existing \cbs improvements}
\changed{
In addition to the PC improvement presented in Section~\ref{sec:improvements}, more \cbs improvements exist.
Some of these include adding heuristics~\cite{felner2018}, disjoint splitting~\cite{li2019}, bypassing a conflict~\cite{boyarski2015b}, symmetry breaking~\cite{li2020} and exploiting similarities between nodes in a single conflict tree~\cite{boyarski2020}.
We can also apply variants of \cbs~\cite{barer2014} that compute (bounded) sub-optimal solutions to \ouralg.
}

\subsection{Extensions to the {\cmapf} Framework}
\label{sec:cmapf_extensions}

\subsubsection{Number and types of collaborating agents}
A rather straightforward generalization of {\cmapf} is to require more than two agents to collaborate on a task. %work together in order to complete a task.
The problem introduced in Section \ref{sec:introduction} motivates this extension: several grasp units may pickup several items for a single transfer unit.
\ouralg can solve this problem with a few minor changes.
However, if the number of agents per task isn't fixed, additional work is required.
Moreover, we may consider agents with different traversal capabilities (e.g., different velocities~\cite{honig2016}), by possibly changing the single-agent planner.

\subsubsection{Other forms of cooperative interaction}
We introduced a %basic 
definition for the {\cmapf} problem, where interaction between agents is expressed via meetings between two types of agents. %, %{\agentAname} and {\agentBname}.
While this interaction is very intuitive, more forms of cooperative interaction can be modeled (for example, temporal constraints).
% For example, we may represent temporal constraints between agents, such as precedence.
% The formulation can be generalized
We may generalize the formulation to include a finite set of possible agent types, and define more complex tasks where each agent type has its dedicated role.% in completing the task.

%It is interesting to note that 
The framework provided by \ouralg might allow to address such general definitions by only adjusting the \emph{cooperation-level} search %i.e., 
(the meetings level in our case). 
Any cooperative planning, which results in inducing goals for an agent, 
%(for the path-level search), 
can be easily plugged in into \ouralg.

\subsubsection{Task assignment and lifelong planning}
In this problem we assume cooperative tasks are pre-assigned to collaborating agents.
However, optimizing the task assignment as well may significantly affect solution quality (as in classical MAPF).
This is extremely relevant for lifelong-planning problems, where agents have to attend to a stream of incoming tasks.
Generalizing the {\cmapf} framework in this direction will advance it even further towards more real-world problems, but introduce significant challenges as well. 

% \ask{TODO: fix references, especially ijcai.org in publisher}
% %%%%%%%%%%%%%%%%%%%%%%%%%%%%%%%%%%%%%%%%%%%%%%%%%%%%%%%%%%%%%%%%%%%%%%%%%%%%%%%%

% \begin{figure}[ht]
%     \centering
%     \includegraphics[width=\columnwidth]{figures/results/num_expansions_SOC.png}
%     % \setlength{\abovecaptionskip}{0pt}
%     \caption{Number of expanded CT nodes. Root and regular nodes are darkly and lightly shaded, respectively.}
%     \label{fig:num_expansions}
%     \vspace{-1.0em}
% \end{figure}
% \vspace{-1.5em}

%%%%%%%%%%%%%%%%%%%%%%%%%%%%%%%%%%%%%%%%%%%%%%%%%%%%%%%%%%%%%%%%%%%%%%%%%%%%%%%%

% % \begin{thebibliography}{99}
% % \end{thebibliography}
% \bibliography{references}

\bibliographystyle{IEEEtran}
\bibliography{IEEEabrv,references}

\begin{thebibliography}{10}
\providecommand{\url}[1]{#1}
\csname url@rmstyle\endcsname
\providecommand{\newblock}{\relax}
\providecommand{\bibinfo}[2]{#2}
\providecommand\BIBentrySTDinterwordspacing{\spaceskip=0pt\relax}
\providecommand\BIBentryALTinterwordstretchfactor{4}
\providecommand\BIBentryALTinterwordspacing{\spaceskip=\fontdimen2\font plus
\BIBentryALTinterwordstretchfactor\fontdimen3\font minus
  \fontdimen4\font\relax}
\providecommand\BIBforeignlanguage[2]{{%
\expandafter\ifx\csname l@#1\endcsname\relax
\typeout{** WARNING: IEEEtran.bst: No hyphenation pattern has been}%
\typeout{** loaded for the language `#1'. Using the pattern for}%
\typeout{** the default language instead.}%
\else
\language=\csname l@#1\endcsname
\fi
#2}}

\bibitem{torreno2017}
A.~Torre{\~{n}}o, E.~Onaindia, A.~Komenda, and M.~Stolba, ``Cooperative
  multi-agent planning: {A} survey,'' \emph{{Computing Research Repository
  (CoRR)}}, vol. abs/1711.09057, 2017.

\bibitem{stern2019}
R.~Stern, N.~R. Sturtevant, A.~Felner, S.~Koenig, H.~Ma, T.~T. Walker, J.~Li,
  D.~Atzmon, L.~Cohen, T.~K.~S. Kumar, R.~Bart{\'{a}}k, and E.~Boyarski,
  ``Multi-agent pathfinding: Definitions, variants, and benchmarks,'' in
  \emph{{Int. Symp. on Combinatorial Search (SOCS)}}, 2019, pp. 151--159.

\bibitem{wurman2008}
P.~R. Wurman, R.~D'Andrea, and M.~Mountz, ``Coordinating hundreds of
  cooperative, autonomous vehicles in warehouses,'' \emph{{Artificial
  Intelligence}}, vol.~29, no.~1, pp. 9--20, 2008.

\bibitem{dresner2008}
K.~Dresner and P.~Stone, ``A multiagent approach to autonomous intersection
  management,'' \emph{{Journal of Artificial Intelligence Research (JAIR)}},
  vol.~31, pp. 591--656, 2008.

\bibitem{vsvancara2019}
J.~{\v{S}}vancara, M.~Vlk, R.~Stern, D.~Atzmon, and R.~Bart{\'a}k, ``Online
  multi-agent pathfinding,'' in \emph{{{AAAI} Conf. on Artificial
  Intelligence}}, vol.~33, 2019, pp. 7732--7739.

\bibitem{honig2016}
W.~H{\"{o}}nig, T.~K.~S. Kumar, L.~Cohen, H.~Ma, H.~Xu, N.~Ayanian, and
  S.~Koenig, ``Multi-agent path finding with kinematic constraints,'' in
  \emph{{Int. Conf. Automated Planning and Scheduling (ICAPS)}}.\hskip 1em plus
  0.5em minus 0.4em\relax {AAAI} Press, 2016, pp. 477--485.

\bibitem{ma2016a}
H.~Ma, S.~Koenig, N.~Ayanian, L.~Cohen, W.~H{\"{o}}nig, T.~K.~S. Kumar,
  T.~Uras, H.~Xu, C.~A. Tovey, and G.~Sharon, ``Overview: Generalizations of
  multi-agent path finding to real-world scenarios,'' \emph{{Computing Research
  Repository (CoRR)}}, vol. abs/1702.05515, 2017.

\bibitem{felner2017}
A.~Felner, R.~Stern, S.~E. Shimony, E.~Boyarski, M.~Goldenberg, G.~Sharon,
  N.~R. Sturtevant, G.~Wagner, and P.~Surynek, ``Search-based optimal solvers
  for the multi-agent pathfinding problem: Summary and challenges,'' in
  \emph{{Int. Symp. on Combinatorial Search (SOCS)}}, 2017, pp. 29--37.

\bibitem{salzman2020}
O.~Salzman and R.~Stern, ``Research challenges and opportunities in multi-agent
  path finding and multi-agent pickup and delivery problems,'' in \emph{{Int.
  Conf. on Autonomous Agents and MultiAgent Systems (AAMAS)}}, 2020, pp.
  1711--1715.

\bibitem{ma2017}
H.~Ma, J.~Li, T.~K.~S. Kumar, and S.~Koenig, ``Lifelong multi-agent path
  finding for online pickup and delivery tasks,'' in \emph{{Int. Conf. on
  Autonomous Agents and MultiAgent Systems (AAMAS)}}, 2017, pp. 837--845.

\bibitem{liu2019}
M.~Liu, H.~Ma, J.~Li, and S.~Koenig, ``Task and path planning for multi-agent
  pickup and delivery,'' in \emph{{Int. Conf. on Autonomous Agents and
  MultiAgent Systems (AAMAS)}}, 2019, pp. 1152--1160.

\bibitem{ma2016b}
H.~Ma, C.~A. Tovey, G.~Sharon, T.~K.~S. Kumar, and S.~Koenig, ``Multi-agent
  path finding with payload transfers and the package-exchange robot-routing
  problem,'' in \emph{{{AAAI} Conf. on Artificial Intelligence}}, 2016, pp.
  3166--3173.

\bibitem{atzmon2020a}
D.~Atzmon, Y.~Zax, E.~Kivity, L.~Avitan, J.~Morag, and A.~Felner,
  ``Generalizing multi-agent path finding for heterogeneous agents,'' in
  \emph{{Int. Symp. on Combinatorial Search (SOCS)}}, 2020, pp. 101--105.

\bibitem{correll2016}
N.~Correll, K.~E. Bekris, D.~Berenson, O.~Brock, A.~Causo, K.~Hauser, K.~Okada,
  A.~Rodriguez, J.~M. Romano, and P.~R. Wurman, ``Analysis and observations
  from the first {Amazon} picking challenge,'' \emph{{IEEE} Transactions on
  Automation Science and Engineering}, vol.~15, no.~1, pp. 172--188, 2016.

\bibitem{shome2020}
R.~Shome, ``Roadmaps for robot motion planning with groups of robots,''
  \emph{Current Robotics Reports}, pp. 1--10, 2021.

\bibitem{murray2020}
C.~C. Murray and R.~Raj, ``The multiple flying sidekicks traveling salesman
  problem: Parcel delivery with multiple drones,'' \emph{Transportation
  Research Part C: Emerging Technologies}, vol. 110, pp. 368--398, 2020.

\bibitem{choudhury2020}
S.~Choudhury, K.~Solovey, M.~J. Kochenderfer, and M.~Pavone, ``Efficient
  large-scale multi-drone delivery using transit networks,'' in \emph{{{IEEE}
  Int. Conf. Robotics and Automation ({ICRA})}}, 2020, pp. 4543--4550.

\bibitem{sharon2015}
G.~Sharon, R.~Stern, A.~Felner, and N.~R. Sturtevant, ``Conflict-based search
  for optimal multi-agent pathfinding,'' \emph{{Artificial Intelligence}}, vol.
  219, pp. 40--66, 2015.

\bibitem{honig2018}
W.~H{\"{o}}nig, S.~Kiesel, A.~Tinka, J.~W. Durham, and N.~Ayanian,
  ``Conflict-based search with optimal task assignment,'' in \emph{{Int. Conf.
  on Autonomous Agents and MultiAgent Systems (AAMAS)}}, 2018, pp. 757--765.

\bibitem{cap2015}
M.~C{\'{a}}p, J.~Vokr{\'{\i}}nek, and A.~Kleiner, ``Complete decentralized
  method for on-line multi-robot trajectory planning in well-formed
  infrastructures,'' in \emph{{Int. Conf. Automated Planning and Scheduling
  (ICAPS)}}, 2015, pp. 324--332.

\bibitem{yu2014}
J.~Yu and D.~Rus, ``Pebble motion on graphs with rotations: Efficient
  feasibility tests and planning algorithms,'' in \emph{{Workshop on the
  Algorithmic Foundations of Robotics (WAFR)}}, ser. Springer Tracts in
  Advanced Robotics, vol. 107, 2014, pp. 729--746.

\bibitem{surynek2020}
P.~Surynek, ``Multi-goal multi-agent path finding via decoupled and integrated
  goal vertex ordering,'' \emph{{Computing Research Repository (CoRR)}}, vol.
  abs/2009.05161, 2020.

\bibitem{sharon2013}
G.~Sharon, R.~Stern, M.~Goldenberg, and A.~Felner, ``The increasing cost tree
  search for optimal multi-agent pathfinding,'' \emph{{Artificial
  Intelligence}}, vol. 195, pp. 470--495, 2013.

\bibitem{gordon2021}
O.~Gordon, Y.~Filmus, and O.~Salzman, ``Revisiting the complexity analysis of
  conflict-based search: New computational techniques and improved bounds,''
  \emph{{Computing Research Repository (CoRR)}}, vol. abs/2104.08759, 2021.

\bibitem{boyarski2015}
E.~Boyarski, A.~Felner, R.~Stern, G.~Sharon, D.~Tolpin, O.~Betzalel, and S.~E.
  Shimony, ``{ICBS:} improved conflict-based search algorithm for multi-agent
  pathfinding,'' in \emph{Int. Joint Conf. on Artificial Intelligence (IJCAI)},
  2015, pp. 740--746.

\bibitem{sturtevant2012}
\BIBentryALTinterwordspacing
N.~Sturtevant, ``Benchmarks for grid-based pathfinding,'' \emph{Transactions on
  Computational Intelligence and AI in Games}, vol.~4, no.~2, pp. 144 -- 148,
  2012. [Online]. Available:
  \url{http://web.cs.du.edu/~sturtevant/papers/benchmarks.pdf}
\BIBentrySTDinterwordspacing

\bibitem{li2021}
J.~Li, D.~Harabor, P.~J. Stuckey, and S.~Koenig, ``Pairwise symmetry reasoning
  for multi-agent path finding search,'' \emph{{Computing Research Repository
  (CoRR)}}, vol. abs/2103.07116, 2021.

\bibitem{atzmon2020b}
D.~Atzmon, J.~Li, A.~Felner, E.~Nachmani, S.~S. Shperberg, N.~Sturtevant, and
  S.~Koenig, ``Multi-directional heuristic search,'' in \emph{Int. Joint Conf.
  on Artificial Intelligence (IJCAI)}, 2020, pp. 4062--4068.

\bibitem{felner2018}
A.~Felner, J.~Li, E.~Boyarski, H.~Ma, L.~Cohen, T.~K.~S. Kumar, and S.~Koenig,
  ``Adding heuristics to conflict-based search for multi-agent path finding,''
  in \emph{{Int. Conf. Automated Planning and Scheduling (ICAPS)}}, 2018, pp.
  83--87.

\bibitem{li2019}
J.~Li, D.~Harabor, P.~J. Stuckey, H.~Ma, and S.~Koenig, ``Disjoint splitting
  for multi-agent path finding with conflict-based search,'' in \emph{{Int.
  Conf. Automated Planning and Scheduling (ICAPS)}}, 2019, pp. 279--283.

\bibitem{boyarski2015b}
E.~Boyarski, A.~Felner, G.~Sharon, and R.~Stern, ``Don't split, try to work it
  out: Bypassing conflicts in multi-agent pathfinding,'' in \emph{{Int. Conf.
  Automated Planning and Scheduling (ICAPS)}}, 2015, pp. 47--51.

\bibitem{li2020}
J.~Li, G.~Gange, D.~Harabor, P.~J. Stuckey, H.~Ma, and S.~Koenig, ``New
  techniques for pairwise symmetry breaking in multi-agent path finding,'' in
  \emph{{Int. Conf. Automated Planning and Scheduling (ICAPS)}}, 2020, pp.
  193--201.

\bibitem{boyarski2020}
E.~Boyarski, A.~Felner, D.~Harabor, P.~J. Stuckey, L.~Cohen, J.~Li, and
  S.~Koenig, ``Iterative-deepening conflict-based search,'' in \emph{Int. Joint
  Conf. on Artificial Intelligence (IJCAI)}, 2020, pp. 4084--4090.

\bibitem{barer2014}
M.~Barer, G.~Sharon, R.~Stern, and A.~Felner, ``Suboptimal variants of the
  conflict-based search algorithm for the multi-agent pathfinding problem,'' in
  \emph{{Int. Symp. on Combinatorial Search (SOCS)}}, 2014.

\end{thebibliography}

\addtolength{\textheight}{-12cm}   % This command serves to balance the column lengths
                                  % on the last page of the document manually. It shortens
                                  % the textheight of the last page by a suitable amount.
                                  % This command does not take effect until the next page
                                  % so it should come on the page before the last. Make
                                  % sure that you do not shorten the textheight too much.

\end{document}